\pdfoutput=1
\PassOptionsToPackage{table}{xcolor}

\documentclass[11pt]{article}

\usepackage[preprint]{acl}

\usepackage{times}
\usepackage{latexsym}
\usepackage{adjustbox}
\usepackage{paralist}
\usepackage{array, booktabs, makecell, multirow}
\usepackage{amssymb}       
\usepackage{pifont}
\usepackage{longtable}
\usepackage{makecell}
\usepackage{todonotes}
\usepackage{enumitem}

\newcommand{\blue}[1]{\textcolor{black}{#1}} 

\newcommand{\cmark}{\textcolor{green}{\checkmark}}
\newcommand{\xmark}{\textcolor{red}{\ding{55}}}

\usepackage[T1]{fontenc}

\usepackage[utf8]{inputenc}

\usepackage{microtype}
\usepackage{geometry}
\usepackage{booktabs}
\usepackage{multirow}
\usepackage{inconsolata}
\usepackage{graphicx}
\usepackage{listings}
\usepackage{scalerel}
\usepackage{cleveref}
\usepackage{hyperref}
\usepackage{titlesec}
\usepackage{fancyvrb}
\usepackage{varwidth}
\usepackage{fontawesome5}

\renewcommand{\arraystretch}{1.2}

\title{\scalerel*{\includegraphics{
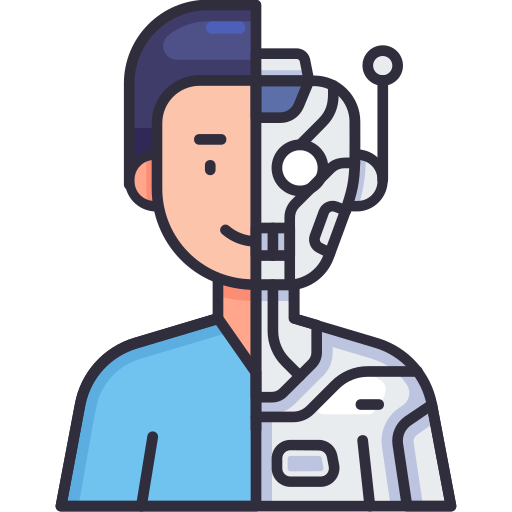}}{\texttt{D}}\hspace{0.1em}\texttt{Droid}: A Resource Suite for AI-Generated Code Detection}

\author{
  Daniil Orel\textsuperscript{1}, 
  Indraneil Paul\textsuperscript{2},  
  Iryna Gurevych\textsuperscript{1,2}, and
  Preslav Nakov\textsuperscript{1} \\ 
  \textsuperscript{1} Mohamed bin Zayed University of Artificial Intelligence (MBZUAI), UAE \\
  \textsuperscript{2} Ubiquitous Knowledge Processing Lab (UKP Lab), Department of Computer Science \\ 
   TU Darmstadt and National Research Center for Applied Cybersecurity ATHENE, Germany \\
   \texttt{\{name.surname\}@mbzuai.ac.ae; \{name.surname\}@tu-darmstadt.de} \\
}

\begin{document}

\setlength{\parskip}{.15pt}
\setlength{\textfloatsep}{3pt minus 2.0pt}
\setlength{\dbltextfloatsep}{7pt minus 2.0pt}
\titlespacing*{\section}{0pt}{1.2ex plus .2ex minus .2ex}{1.2ex plus .2ex}
\titlespacing*{\subsection}{0pt}{1.2ex plus .2ex minus .2ex}{1.2ex plus .2ex}

\maketitle
\begin{abstract}
In this work, we compile \textbf{\texttt{DroidCollection}}\footnote{\href{https://huggingface.co/collections/project-droid/droid-683360d8b008214a4273099a}{\scalerel*{\includegraphics{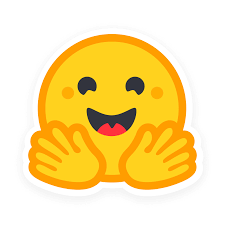}}{\texttt{|}}\hspace{0.2em}\texttt{https://huggingface.co/collections/ project-droid/droid-683360d8b008214a4273099a}}}
\footnote{\href{https://tudatalib.ulb.tu-darmstadt.de/items/ebc68cfb-186e-4303-bd46-cbd015af2045}{\scalerel*{\includegraphics{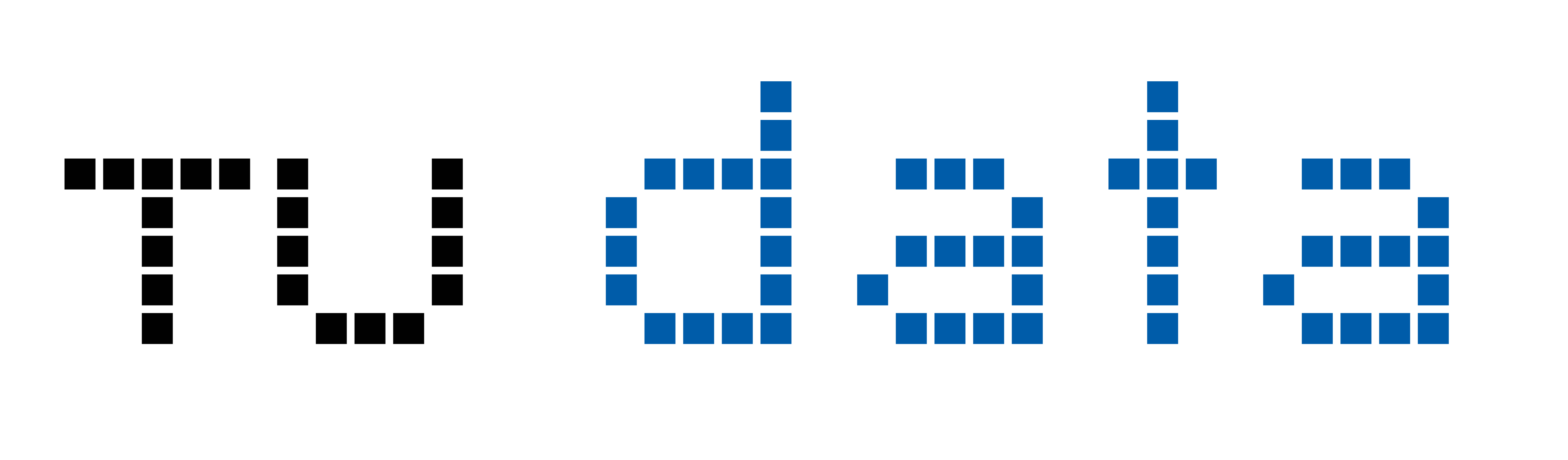}}{\texttt{|}}\hspace{0.2em}\texttt{https://tudatalib.ulb.tu-darmstadt.de/ items/ebc68cfb-186e-4303-bd46-cbd015af2045}}}, the most extensive open data suite for training and evaluating machine-generated code detectors, comprising over a million code samples, seven programming languages, outputs from 43 coding models, and over three real-world coding domains. Alongside fully AI-generated samples, our collection includes human-AI co-authored code, as well as adversarial samples explicitly crafted to evade detection. Subsequently, we develop \textbf{\texttt{DroidDetect}}, a suite of encoder-only detectors trained using a multi-task objective over \texttt{DroidCollection}. Our experiments show that existing detectors' performance fails to generalise to diverse coding domains and programming languages outside of their narrow training data. Additionally, we demonstrate that while most detectors are easily compromised by humanising the output distributions using superficial prompting and alignment approaches, this problem can be easily amended by training on a small amount of adversarial data. Finally, we demonstrate the effectiveness of metric learning and uncertainty-based resampling as means to enhance detector training on possibly noisy distributions.

\end{abstract}

\section{Introduction}
\label{sec:Intro}

In recent years, language models (LMs) for code generation (Code-LMs) have become a near-indispensable accessory in a developer's toolbox. Their enhancement of productivity has proliferated into most of the software development lifecycle, including automating unit test generation~\citep{DBLP:conf/iclr/JainSR25}, code infilling~\citep{DBLP:journals/corr/abs-2207-14255}, predicting build errors, and code refactoring, \textit{inter alia}, propelling their broad adoption in production~\citep{DBLP:conf/sigsoft/DunayCTTRCAGMMT24, DBLP:conf/icse/FrommgenA24, DBLP:journals/pacmse/MuraliMABCGFNR24}. However, the code authoring and refinement abilities of these models present issues with respect to domains where the human authorship of the generated artefacts is paramount and the consequences of limited human supervision are of concern.

Despite the well-documented productivity benefits of using AI assistance for knowledge workers~\citep{DBLP:journals/pacmhci/WeberBSM24,DBLP:conf/nips/LiSCLZ23}, there exists a wide range of scenarios where ensuring the human authorship of artefacts is vital, resulting in the need for robust detectors of machine-assisted code. For instance, in academia, students' reliance on LMs for assignments undermines educational integrity, with professors unable to detect the authorship of submissions and grading AI-generated coding practices~\citep{DBLP:conf/aaai/KoikeKO24}. Similarly, conducting technical hiring fairly and human code annotation studies accurately requires the ability to ensure that the submitted artefacts are authentically human-authored~\citep {DBLP:journals/corr/abs-2306-07899}.

The subtle failure patterns in the outputs of code LMs imply the need for strong detection mechanisms as part of the workflow in order to safeguard against unforeseen side effects. For instance, machine-generated code can introduce serious vulnerabilities (\emph{e.g.,} insecure logic, hidden backdoors, or injection flaws), which can jeopardise software reliability~\citep{bukhari2024issues} and data security~\citep{DBLP:journals/cacm/PearceATDK25}. It can also facilitate obfuscation, producing code that is harder to parse~\citep{copilot_obstacles}: this can hide malicious functionality and complicate debugging~\citep{DBLP:journals/corr/abs-2502-02368}. These weaknesses can amplify over time, creating a dangerous feedback loop where (possibly deficient) AI-generated code enters public repositories and is leveraged for subsequent training runs, thus increasing the risk of degraded data quality~\citep{ji2024cybersecurity} or, even worse, collapsing models~\citep{DBLP:journals/nature/ShumailovSZPAG24}.

Despite the increasing interest in detecting AI-generated code, most current work has notable limitations. Existing work usually covers fewer than three programming languages~\citep{gptsensor} and focuses on a narrow set of API-based code generators~\citep{DBLP:journals/corr/abs-2310-05103}. Moreover, detectors typically address the problem as a binary classification task: machine-generated vs. human-written code~\citep{DBLP:conf/coling/JawaharASL20}. This ignores common hybrid operating modes where code is co-authored by humans and LMs or adversarial scenarios where models are prompted or tuned to evade detection~\citep{DBLP:journals/corr/abs-2408-04284}.

Our work addresses these limitations with a comprehensive and scalable approach to AI-generated code detection. Our contributions are as follows:
\begin{itemize}[leftmargin=*]

\item We compile and open-source \texttt{DroidCollection}, an extensive suite of multi-way classification data for training and evaluating AI-generated code detectors. \texttt{DroidCollection} contains over 1 million instances sampled from 43 LMs (spanning 11 model families), 7 programming languages, and multiple coding domains.

\item We propose a novel task: detection of code generated by adversarially trained LMs, which mimics intentional obfuscation and evasion behaviours. To this end, we compile and release \texttt{DroidCollection-Pref}, a preference corpus of 157k response pairs \blue{designed to encourage language models to produce responses that closely resemble those of humans.}

\item We open-source \texttt{Dro\-id\-De\-tect-Ba\-se} and \texttt{Dro\-id\-De\-tect-Lar\-ge}, two state\--of\--the\--art AI-generated code detectors fine-tuned from Mo\-dern\-BERT~\citep{modernbert} Base (149M), and Large (396M) models, respectively, using \texttt{Droid\-Col\-lec\-tion}.

\item We conduct extensive out-of-distribution performance analysis across languages, coding domains and detection settings. Our evaluation results demonstrate that there is positive transfer across related programming languages~\citep{DBLP:conf/hapoc/Martini15} and across domains. We also find that most existing models struggle when tasked with detecting machine-refined code and are almost entirely unusable against adversarially humanised model-generated code. However, we show that this can be rectified by incorporating modest amounts of such data during training.

\end{itemize}

\section{Related Work}
\label{sec:Related}

We briefly outline three relevant lines of existing work: \textbf{1)} AI-generated text detection, \textbf{2)} AI-generated code detection, and \textbf{3)} adversarial evasion of AI-generated content detectors.

\subsection{AI-Generated Text Detection}

Early research on synthetic data detection has focused on detecting AI-generated text in specific, fundamental tasks such as question answering~\citep{hc3}, translation, summarisation and paraphrasing~\citep{hc3_plus}. Major early contributions to creating comprehensive benchmarks include M4~\citep{m4}, which introduced a multilingual, multi-generator and multi-domain benchmark consisting of 122,000 human-written and machine-generated texts. MULTITuDE~\citep{multitude} featured a multilingual dataset with over 70,000 samples of AI and human-written texts across 11 languages. Additionally, MAGE~\citep{mage} concentrated on English-only scenarios, but emphasised evaluating model robustness by testing across eight distinct out-of-domain settings to simulate real-world scenarios.
The advancement of this field has been further stimulated by numerous competitions and shared tasks dedicated to AI-generated text detection, including RuATD~\citep{ruatd}, a shared task at COLING'2025~\citep{wang-etal-2025-genai}, a shared task at ALTA~\citep{alta}, and DagPap~\citep{dagap}. 

Tools such as MOSS~\citep{moss_copilot} have shown some effectiveness in identifying AI-generated code, since their style is out of the ordinary distributions of student solutions. However, \citet{detectors} and \citet{aigc} have shown that detectors such as GPT-Zero often fail when applied to code rather than text. This critical observation, backed up by our experiments in \Cref{sec:ood}, highlights the inadequacy of directly porting generic text-based models to the code domain and strongly motivates the creation of code-specific detection strategies and specialised datasets. Our work responds to this need by providing a large-scale, multifaceted suite specifically curated for AI-generated code, designed to foster the development and rigorous testing of detection techniques attuned to the unique characteristics of programming languages and AI-generated software.

\subsection{AI-Generated Code Detection}

Early attempts at AI-generated code detection using decision tree learning methods, such as \citet{whodunit} and \citet{DBLP:conf/issre/LiHFZL23}, demonstrated that code-level statistics (e.g., number of lines, Abstract Syntax Tree (AST) depth, identifier length) can serve as reliable indicators of authorship. However, more sophisticated detection requires a more involved feature engineering, which is best performed using deep learning methods~\citep{DBLP:conf/nips/TulchinskiiKKCN23}. Thus, recent efforts have mainly focused on training text-based LMs to detect AI-generated code. 
A common approach in existing work, such as GPTSniffer \citep{NGUYEN2024112059} and GPT-Sensor \citep{gptsensor} is to extract human-written functions from the Code\-Search\-Net dataset~\citep{codesearchnet} and then to prepare machine-generated counterparts to them using ChatGPT. 
Although similar in their dataset construction, these two works differ in modelling.
GPTSniffer uses a multi-class classification loss, whereas GPT-Sensor applies a cosine similarity-based loss to better separate the embeddings of AI-generated and human-written code, aiming at learning more discriminative representations.

To address the overdependence on CodeSearchNet in prior work, \citet{orel2025codetm4detectingmachinegeneratedcode} source additional code from LeetCode and CodeForces. They evaluated a wide range of locally deployable LMs as code generators and provided a systematic analysis of out-of-distribution (OOD) detection performance across different settings.
Importantly, they go beyond binary classification by introducing more nuanced scenarios, such as collaborative (or hybrid) settings where LMs complete or rewrite human-written programs. 

CodeMirage~\cite{codemirage}, another benchmark released concurrently with ours, also includes rewritten code. However, these works lack diversity in terms of generators: CoDet-M4 covers only 5 generators, while CodeMirage covers 10.
Furthermore, they overlook the importance of diverse sampling strategies in the detection of machine-generated codes and do not consider more adversarial settings where the model-generated code is artificially humanised.
Our work builds upon and extends the progress of previous works by further increasing the scale and diversity: we incorporate three distinct domains -- \blue{competitive programming solutions (e.g., LeetCode), open-source GitHub repositories, and code from research papers} -- utilise 43 generative models, and cover seven programming languages. 
Notably, unlike Co\-det-M4 or Code\-Mirage, our dataset is the first in this domain to systematically integrate diverse sampling strategies using varied generation settings and synthetic data generation scenarios.

\subsection{Adversarial Evasion of AI-Generated Content Detectors}

Although specialized detectors for AI-generated code can be effective against honest actors, their straightforward training on machine-generated and machine-refined data makes them vulnerable to adversarially perturbed or humanised text, modified to evade detection~\citep{DBLP:conf/ccs/0001SC0024,DBLP:journals/corr/abs-2501-03437}. Currently, RAID~\citep{raid}, one of the most extensive benchmarks in AI-generated text detection, is notable in being one of the few efforts exploring adversarial detection settings with various attack methods such as paraphrasing and synonym substitution. Our work in AI-generated code detection builds upon this important aspect. We extend this focus to the code \blue{modality} by incorporating a diverse set of adversarial attack scenarios specifically engineered to challenge detectors.
Moreover, we move beyond the language manipulations considered by RAID to address the possibilities of adversarial training using targeted mining of paired preference data and a dedicated collection of adversarial prompting, which are all aspects that are vital for assessing detector robustness under more challenging conditions.

\section{The \texttt{DroidCollection} Corpus}
\label{sec:Dataset}

\begin{table*}[htbp]
    \small
    \centering
    \scalebox{0.75}{
    \begin{tabular}{rlcccccc}
    \toprule
    \multirow{2}{*}{\textbf{\texttt{Name}}} & 
    \multirow{2}{*}{\textbf{\texttt{Size}}} & 
    \textbf{\texttt{Supported}} &
    \textbf{\texttt{No. of}} & 
    \textbf{\texttt{Supported}} &  \textbf{\texttt{Varied}} & \textbf{\texttt{Machine}} & \textbf{\texttt{Adversarially}} \\
    
     &  & \textbf{\texttt{Domains}} & \textbf{\texttt{Models}} & \textbf{\texttt{Languages}} & \textbf{\texttt{Sampling}} & \textbf{\texttt{Refined Data}} & \textbf{\texttt{Humanized Data}} \\
        
    \midrule
    \texttt{GPT-Sniffer}~\citep{NGUYEN2024112059} & \texttt{7.4k} & \texttt{1} & \texttt{1} & \texttt{2} & \xmark & \xmark & \xmark \\
    \midrule
    \texttt{CodeGPTSensor}~\citep{gptsensor} & \texttt{1.1M} & \texttt{1} & \texttt{1} & \texttt{2} & \xmark & \xmark & \xmark \\
    \midrule
    \texttt{Whodunit}~\citep{whodunit} & \texttt{1.6k} & \texttt{1} & \texttt{1} & \texttt{1} & \xmark & \xmark & \xmark \\
    \midrule
    \texttt{CoDet-M4}~\citep{orel2025codetm4detectingmachinegeneratedcode} & \texttt{501K} & \texttt{2} & \texttt{5} & \texttt{3} & \xmark & \cmark & \xmark \\
     \midrule
    \texttt{CodeMirage}~\citep{codemirage} & \texttt{210k} & \texttt{1} & \texttt{10} & \texttt{10} & \xmark & \cmark & \xmark \\
    \midrule
    \textbf{\texttt{DroidCollection}} & \texttt{\textbf{1.06M}} & \texttt{\textbf{3}} & \texttt{\textbf{43}} & \texttt{\textbf{7}} & \cmark & \cmark & \cmark \\
    \bottomrule
    
    \end{tabular}}
    \caption{Comparison of \texttt{DroidCollection} to other AI-generated code detection datasets \blue{shows broader domain and model coverage, as well as unique inclusion of varied sampling strategies and adversarially humanized data}.}

    \label{tab:dataset_comparison}
\end{table*}
\blue{We compare our dataset with existing ones in \Cref{tab:dataset_comparison}, which shows that our dataset is not only among the largest to date, but also captures a broader range of variations. Additional details about key characteristics are provided in \Cref{appx:stats}.}

In this section, we detail the curation of the human-generated, machine-generated, and machine-refined splits of \texttt{DroidCollection}. The adversarially humanised data collection is deferred to \Cref{sec:adversarial}.

\subsection{Human-Authored Code Acquisition}

To build the dataset, we collected human-written samples from multiple sources, covering C++/C, C\#, Go, Java, JavaScript and Python languages. 
Then, we generated code, using base and instruction-tuned LMs from 11 model release families, namely: Llama~\citep{llama3.1}, CodeLlama, GPT-4o, Qwen~\citep{qwen}, IBM Granite~\citep{mishra2024granitecodemodelsfamily}, Yi~\citep{Young2024YiOF}, DeepSeek~\citep{guo2024deepseekcoderlargelanguagemodel}, Phi~\citep{abdin2024phi4technicalreport}, Gemma~\citep{team2024gemma}, Mistral~\citep{mistral2025}, Starcoder~\citep{li2023starcodersourceyou}. 
The list of generators per model family is detailed in \Cref{appx:model_names}. Our dataset covers three domains: general use code, algorithmic problems, and research/data-science code.

\paragraph{General Use Code} \blue{This type of code is} normally deployed for disparate use cases such as web serving, firmware, game engines, etc. These are largely hosted on GitHub, and mainly obtained from StarcoderData~\citep{li2023starcodersourceyou}, and The Vault~\citep{thevault} datasets.

\paragraph{Algorithmic Problems} This category contains code solutions to competitive programming problems -- \blue{algorithmic challenges designed to test problem-solving skills, commonly featured in competitions}. These are retrieved from multiple sources such as TACO~\citep{taco}, CodeNet~\citep{codenet} (mainly AtCoder\footnote{\url{https://atcoder.jp/}} and AIZU\footnote{\url{https://onlinejudge.u-aizu.ac.jp/home}} platforms), LeetCode and CodeForces, from the work of ~\citet{orel2025codetm4detectingmachinegeneratedcode}. Its primary distinguishing feature is its tendency to contain simple and self-contained routines.

\paragraph{Research Code} This \blue{data subset} is sourced from code repositories accompanying research papers, \blue{collected in} ObscuraCoder~\citep{obscura}. Additionally, we augment this \blue{data} with mathematical and data science code \blue{from work of} ~\citet{DBLP:conf/iclr/LuZ0RSPZL25}. This subset is characterised by the lack of modularity and the over-representation of procedural code.

\subsection{AI-Authored Code Generation}
\label{sec:generation}
 

\paragraph{Generation via Inverse Instruction}
Since the data from sources such as CodeNet and StarcoderData do not contain any instructions, we decided to apply inverse instructions to transform code from these datasets into instructions, which can be used to prompt LMs. 
In our case, the method of preparing inverse instructions was similar to that described in InverseCoder~\citep{inversecoder}: we passed the code snippets to an LM, asking it to build a summary, and a step-by-step instruction that can be given to an LM to generate a similar code.
The main difference between our approach and that of InverseCoder is that we \blue{sought} to minimise the costs of the generation and did not split the summarisation and instruction generation into separate LM calls. However, in cases where a summary could be extracted from the response but the instruction could not, we used the summary to re-generate the instruction. 
This experiment with details about the prompts and the models we used is illustrated in \Cref{appx:inverse}.
This type of generation allows us to cover a wide range of prompts, simulating a diversity of user-LM interactions, which is common in the real world.

\paragraph{Generation Based on Comments}
Some of the data sources used in our study provide docstrings (The Vault Class and Function) or comments (The Vault Inline) that describe the given code. 
In this case, we mainly used base models, which were prompted with the first line of code and the docstring or comment for generation.
Instruct models were given only the docstring and a task to implement the desired class or function.

\paragraph{Generation Based on a Task}
The examples from platforms with algorithmic problems mainly come with a precise task description or a problem statement. 
In this case, we only used the description to prompt the LMs for generation.

\paragraph{Unconditional Synthetic Data}

\blue{This data is not conditioned on prior human generations}. 
\blue{The rationale behind this is that } the machine-generated data used to train the majority of AI-generated content detectors is acquired in a biased manner. 
It usually involves completing incomplete human-written samples or responding to prompts conditioned on existing human generations. This bias, though rather subtle, leads to a situation where detectors are only exposed to the kinds of synthetic data that are easiest for the models to learn~\citep{DBLP:journals/corr/abs-2412-02595}. Hence, we seek to obtain synthetic data that is not conditioned on prior human generations. Following prior work\footnote{\url{https://huggingface.co/blog/cosmopedia}}, we create synthetic user profiles on which we condition coding tasks and, in turn, the final generated code.

To explore how large language models can simulate the behaviour profiles of real programmers, we took inspiration from the PersonaHub dataset~\citep{persona}. We first generated a diverse set of programmer profiles, and then used an LM to create programming tasks that can typically be performed by programmers of such types.
These tasks, along with their corresponding descriptions, termed \texttt{DroidCollection-Personas}, were then used to generate code samples. More details about the \texttt{DroidCollection-Personas} are outlined in the \Cref{appx:persona}.

\subsection{Machine-Refined Data}
In practice, purely AI-generated code is rare. 
Developers typically collaborate with LMs, starting with human-written code and asking the model to modify or extend it. 
This makes binary classification (human vs. machine) insufficient for real-world scenarios.
Instead, introducing a third class to capture human-LLM collaboration, as proposed by \citet{orel2025codetm4detectingmachinegeneratedcode}, offers a more realistic and useful approach.

To generate \blue{such samples}, we designed three scenarios: \emph{(i)} \emph{Human-to-LLM continuation}: A human initiates the code, and the LM completes it. We simulated this by preserving the first $N\%$ of the code lines (where $N$ ranges from 10\% to 85\%) and asking the model to complete the rest. \emph{(ii)} \emph{Gap filling}: The model is given the beginning and the end of a human-written code snippet and is asked to generate the missing part in the middle. The amount of preserved code follows a similar proportion as in the first scenario. \emph{(iii)} \emph{Code rewriting}: The LM is asked to rewrite human-authored code, either with no specific prompt or with an instruction to optimise it.

\subsection{Adversarial Samples}
\label{sec:adversarial}
\blue{With the development of post-training techniques such as PPO~\citep{ppo}, DPO~\citep{dpo}, and GRPO~
\citep{grpo}, it has become possible to set up the training in adversarial ways that enable LMs to evade AI-generated code detectors. Prior work by \citet{red_teaming} and \citet{can_ai_generated_text_be_detected} has shown that LM-generated content detectors are vulnerable to adversarial attacks and spoofing. This motivates the inclusion of adversarial samples in \texttt{DroidCollection} to improve model robustness.}

\blue{To this end, we introduce two types of adversarial attacks: prompt-based attacks and preference-tuning-based attacks. In the prompt-based setting, we construct adversarial prompts by instructing the model to ``write like a human'' in multiple ways, relying on the models' parametric knowledge of how to produce outputs that mimic human-authored code and thus challenge detection systems. In the preference-tuning-based setting, we curate \texttt{DroidCollection-Pref}, a dataset of 157K paired examples consisting of human-written and LM-generated code responses to the same prompt. Using \texttt{Droid\-Collection\--Pref}, we train LMs with up to 9B parameters --including LLaMA, Qwen, and Yi--, using LoRA~\citep{hu2022lora} with rank 128 and DPO for two epochs. These models' output distributions are, in effect, steered towards preferring human-like code, making them less likely to contain the stylistic \blue{features} of machine-generated code. Once trained, the models are used to generate new ``machine-humanised'' code samples. We filter their \blue{outputs} as in \Cref{sec:filtering} to keep only high-quality adversarial examples. \blue{Finally}, we obtained a nearly 1:1 ratio of prompt-based vs. preference-tuning adversarial attacks. }

\subsection{Varying Decoding Strategies}


\begin{table}[ht]
\centering
\renewcommand{\arraystretch}{1.3}
\small
\scalebox{0.75}{
\begin{tabular}{lcc}
\toprule
\textbf{Strategy} & \textbf{\texttt{Attribute}} & \textbf{\texttt{Range}}\\ 
\midrule
\texttt{Greedy Decoding} & — & — \\
\multirow{3}{*}{Sampling} & \texttt{Temperature} & \texttt{\{0.1, 0.4, 0.7, 1.0, 1.5, 2.0\}} \\
 & \texttt{Top-k} & \texttt{\{10, 50, 100\}} \\
 & \texttt{Top-p} & \texttt{\{1.0, 0.95, 0.9, 0.8\}} \\
\texttt{Beam Search} & \texttt{Beam Width} & \texttt{\{2, 4, 6, 8\}} \\
\bottomrule
\end{tabular}}
\caption{Decoding settings used for the AI-generated, AI-refined, and AI-humanised splits of \texttt{DroidCollection}.}
\label{tab:decoding_strategies}
\end{table}

It was shown by \citet{sampling} that after greedy decoding, it was easier to detect AI-generated texts compared to when using other decoding techniques.
\blue{To capture this challenging behaviour and reflect the diversity of generative systems}, we experimented with various decoding strategies, as shown in \Cref{tab:decoding_strategies}.

\subsection{Data Filtering}
\label{sec:filtering}
To ensure the quality of our \texttt{DroidCollection} dataset, we applied a series of filtering criteria, commonly used in other code-related works~\citep{ lozhkov2024starcoder, li2023starcodersourceyou,obscura}. 
First, we removed code samples that could not be successfully parsed into an \blue{abstract-syntax tree (AST)}. 
We also filtered samples based on \blue{the} AST depth, keeping only \blue{those} with \blue{the} depth between 2 and 31, to avoid too simple or too complex codes.
We restricted each sample's maximum line length to be between 12 and 400 characters, and the average line length to fall between 5 and 140 characters, and used only samples with between 6 and 300 lines of code. 
Moreover, we filtered samples according to the fraction of alphanumeric characters, retaining only those between 0.2 and 0.75, to avoid the usage of configs and auto-generated files. 
To ensure English documentation, we used the Lingua language detector\footnote{\href{https://github.com/pemistahl/lingua}{GitHub: pemistahl/lingua-py}} and retained only samples where the docstrings showed greater than 99\% confidence of being English. 
Finally, we removed duplicate or near-duplicate samples; for this, we used MinHash~\citep{minhash}, with a shingle size of 8 and a similarity threshold of 0.8.

\blue{It should be acknowledged that we did not filter out the human-written codes which were sourced after the coding copilots became popular. It means, that there is a possibility that among human written codes there could be samples, created with the help of LLMs. This potential issue is tackled in \Cref{sec:training}}.


\begin{table*}[htbp]
\centering
\scalebox{0.55}{
\begin{tabular}{ll*{8}{r}}
\toprule
 & \texttt{\textbf{Model}} 
 & \multicolumn{4}{c}{\texttt{\textbf{2-Class}}} 
 & \multicolumn{4}{c}{\texttt{\textbf{3-Class}}} \\
\cmidrule(lr){3-6} \cmidrule(lr){7-10}
 &  
 & \textbf{\texttt{General}} & \textbf{\texttt{Algorithmic}} & \textbf{\texttt{Research/DS}} & \texttt{\textbf{Avg.}}
 & \textbf{\texttt{General}} & \textbf{\texttt{Algorithmic}} & \textbf{\texttt{Research/DS}} & \texttt{\textbf{Avg.}}\\
\midrule

\multirow{5}{*}{\texttt{Zero-Shot Baselines}} 
& \texttt{Fast-DetectGPT}~\citep{fastdetect}    & \texttt{75.07} & \texttt{63.05} & \texttt{65.43} & \texttt{67.85} & \texttt{66.43} & \texttt{62.90} & \texttt{64.30} & \texttt{64.54} \\
& \texttt{CoDet-M4}~\citep{orel2025codetm4detectingmachinegeneratedcode}          & \texttt{53.41} & \texttt{44.63} & \texttt{65.43} & \texttt{54.49} & \texttt{41.90} & \texttt{46.06} & \texttt{55.43} & \texttt{47.80}  \\
& \texttt{M4}~\citep{m4}                & \texttt{50.17} & \texttt{57.91} & \texttt{44.67} & \texttt{50.92} & \texttt{56.46} & \texttt{58.13} & \texttt{51.21} & \texttt{55.27} \\
& \texttt{GPTSniffer}~\citep{NGUYEN2024112059}         & \texttt{54.25} & \texttt{36.85} & \texttt{32.10} & \texttt{41.07} & \texttt{45.22} & \texttt{31.75} & \texttt{39.88} & \texttt{38.95}\\
& \texttt{GPTZero}            & \texttt{54.05} & \texttt{71.96} & \texttt{44.73} & \texttt{56.91} & \texttt{50.56} & \texttt{66.13} & \texttt{30.62} & \texttt{49.10} \\
\midrule

\multirow{3}{*}{\texttt{OOD Evaluation}}
& \texttt{DroidDetect\textsubscript{CLS}-Base\textsubscript{General}}      & \texttt{99.30} & \texttt{53.73} & \texttt{76.46} & \texttt{76.50} & \texttt{93.05} & \texttt{46.22} & \texttt{76.99} & \texttt{72.09} \\
& \texttt{DroidDetect\textsubscript{CLS}-Base\textsubscript{Algorithmic}}  & \texttt{49.63} & \texttt{98.26} & \texttt{60.78} & \texttt{69.56} & \texttt{47.86} & \texttt{92.84} & \texttt{56.58} & \texttt{65.76} \\
& \texttt{DroidDetect\textsubscript{CLS}-Base\textsubscript{Research/DS}}     & \texttt{47.01} & \texttt{48.02} & \texttt{72.55} & \texttt{55.86} & \texttt{47.86} & \texttt{38.73} & \texttt{59.97} & \texttt{48.85} \\
\midrule

\multirow{2}{*}{\texttt{Fine-Tuned Baselines}}
& \texttt{GCN}          & \texttt{78.57} & \texttt{60.61} & \texttt{67.79} & \texttt{68.99} & \texttt{56.85} & \texttt{46.91} & \texttt{51.13} & \texttt{51.63} \\
& \texttt{CatBoost}     & \texttt{89.69} & \texttt{87.29} & \texttt{77.21} & \texttt{84.73} & \texttt{78.86} & \texttt{74.01} & \texttt{64.07} & \texttt{72.31} \\
\midrule

\multirow{5}{*}{\texttt{Full Training}}
& \texttt{M4\textsubscript{FT}} & \texttt{92.99} & \texttt{89.36} & \texttt{73.99} & \texttt{85.45} & \texttt{80.98} & \texttt{80.72} & \texttt{58.80} & \texttt{73.50} \\
& \texttt{GPT-Sniffer\textsubscript{FT}} & \texttt{97.72} & \texttt{96.52} & \texttt{80.46} & \texttt{91.56} & \texttt{89.42} & \texttt{88.12} & \texttt{70.72} & \texttt{82.75} \\
& \texttt{CoDet-M4\textsubscript{FT}} & \texttt{98.89} & \texttt{98.23} & \texttt{83.77} & \texttt{93.63} & \texttt{85.46} & \texttt{90.41} & \texttt{73.88} & \texttt{83.25} \\
& \texttt{DroidDetect\textsubscript{CLS}-Base}    & \texttt{99.22}& \texttt{98.22} & \texttt{87.57} & \texttt{95.00} & \texttt{92.78} & \texttt{93.05} & \texttt{74.46} & \texttt{86.76}  \\
& \texttt{DroidDetect\textsubscript{CLS}-Large}    & \texttt{\textbf{99.38}} & \texttt{\textbf{98.39}} & \texttt{\textbf{93.24}} & \texttt{\textbf{97.00}} & \texttt{\textbf{93.08}} & \texttt{\textbf{92.86}} & \texttt{\textbf{80.42}} & \texttt{\textbf{88.78}}  \\
\bottomrule
\end{tabular}}
\caption{Comparison of models in 2-Class (human- vs machine-generated) and 3-Class (human- vs machine-generated vs machine-refined) classification setups across programming languages in terms of weighted F1-score. In the OOD section, we show models trained on each domain individually. \texttt{FT} subscript in Full Training section means that this model was fine-tuned on \texttt{DroidCollection}
The best results are shown in \textbf{bold}.}
\label{tab:domain_eval}
\end{table*}

\section{Detection Experiments}
\label{sec:ood}
\subsection{Experimental Setup}
We first evaluate a range of detectors to assess the strengths and limitations of existing AI-generated code detection models. We include several off-the-shelf detectors as our zero-shot baselines and models trained on \texttt{DroidCollection} as fine-tuned baselines.

The baselines include models widely used in related works:
\emph{(i)} \textbf{GPT-Sniffer}\citep{NGUYEN2024112059}, a CodeBERT-based binary classifier fo AI-Generated code detection;
\emph{(ii)} \textbf{CoDet-M4}\citep{orel2025codetm4detectingmachinegeneratedcode}, a UnixCoder model trained on outputs from multiple code generators;
\emph{(iii)} \textbf{M4 classifier}\citep{m4}, for general AI-generated text detection;
\emph{(iv)} \textbf{Fast-DetectGPT}\citep{fastdetect}, a distribution-based zero-shot detector; and
\emph{(v)} \textbf{GPT-Zero}\footnote{\url{https://gptzero.me/}}, an API-based detector (as this API is paid, we evaluated it on a representative sample of 500 code snippets for each label-language and label-domain pair). Since most of these baselines use binary classification, we map our ternary labels (human-written, AI-generated, AI-refined) into binary targets for fair comparison.

Additionally, we train other models directly on our dataset using a multi-class objective:
\emph{(i)} a simple \textbf{GCN}\citep{DBLP:conf/iclr/KipfW17};
\emph{(ii)} a \textbf{CatBoost classifier}\citep{catboost}, following procedures similar to the \emph{Whodunit} paper; and
\emph{(iii)} two encoder-only transformers, ModernBERT-Base and ModernBERT-Large~\citep{modernbert}, denoted as \texttt{DroidDetect\textsubscript{CLS}-Base} and \texttt{DroidDetect\textsubscript{CLS}-Large}.
Full details are provided in \Cref{appx:gnn,appx:catboost,sec:training}.

\subsection{RQ1: What is the value of extensive data collection for training robust detectors?}
\Cref{tab:dataset_comparison} highlights a key limitation of existing datasets: they often lack \emph{diversity}. \Cref{tab:domain_eval,tab:lang_eval} show that this limitation significantly impacts detector performance in realistic settings. Baseline detectors - the ones illustrated in Zero-Shot Baselines section of the tables - underperform on our test split compared to even simple fine-tuned baselines like GCN and CatBoost.
Among the baselines evaluated, zero-shot Fast-DetectGPT consistently yields strong performance across both languages and domains, outperforming all other baselines.
In contrast, pre-trained models usually perform well only on languages and domains that are closely aligned with their original training data. This highlights the limitations of previously collected datasets, which do not cover the diversity of generations in \texttt{DroidCollection}, and hence are far from being useful in real-life scenarios.

The Full Training section of tables contains the variations of pre\--trained models from Zero\--Shot Baselines, but fine\--tuned on \texttt{Droid\-Col\-lec\-ti\-on}. 
It is clearly seen, that after training on \texttt{Droid\-Col\-lec\-ti\-on} data, the performance of the models significantly increases. They start outperforming GCN and CatBoost, but still lag behind \texttt{Droid\-Detect\textsubscript{CLS}} models.

\texttt{DroidDetect\textsubscript{CLS}} models trained on our extensive training split consistently outperform all baselines, achieving near-ideal scores in both binary and ternary tasks. Notably, the larger backbone (\texttt{DroidDetect\textsubscript{CLS}-Large}) dominates across all settings, demonstrating that both model size and diverse training data are crucial for high classification performance.

\subsection{RQ2: How well do models generalise in OOD settings?}
We evaluated the \texttt{Droid\-De\-tect\textsubscript{CLS}-Base}
backbone in OOD settings under \emph{language shift} and \emph{domain shift} conditions: training it on a single programming language or domain. Comparing the multi-domain and the multi-lingual performance of the baselines to our backbone models trained in such restricted conditions allows us \emph{(i)} to uncover possible shortcomings in the training data curation of popular baseline models, as they can be compared head-to-head to both the split-specific and fully-trained variants of our backbone, and \emph{(ii)} to assess the inherent ease with which models are able to transfer along these settings, by-proxy outlining the value of extensive training data curation. 
We selected the base version of the backbone for this restricted training scenario since it was comparable to most of the chosen baselines in terms of size.

\Cref{tab:lang_eval} shows that, under restricted training conditions, models tend to generalise better to syntactically similar languages. For example, a model trained on C/C++ performs reasonably well on C\# and Java. However, for typologically isolated languages such as Python or Java\-Script, all models not trained specifically for it tend to struggle in this setting.
\Cref{tab:domain_eval} illustrates that the models trained on a single domain have a high discrepancy in scores for other domains. For example, as can be seen in the OOD Evaluation section, the classifiers trained on algorithmic problems suffer with the general codes, and show comparable low performance on Research/DS codes, and vice versa.

\subsection{RQ3: How robust are models to adversarial samples?}

Finally, we test adversarial robustness using challenging samples designed to evade detection (\Cref{tab:adversarial_eval}). Here, many baseline detectors struggle: for instance, GPT-Zero achieves only 0.10 recall. M4 and CoDet-M4 show higher recall on adversarial samples but also have high false positives, misclassifying human-written code.
Interestingly, fine-tuning on \texttt{DroidCollection} also improves the recall of baselines for human-written cases. Remarkably, the recall of M4 on adversarial cases drops from 0.73 to 0.67, while the recall on human-written cases increases significantly from 0.40 to 0.91. A similar pattern is observed for CoDet-M4, while for GPT-Sniffer, both scores increase.

In contrast, \texttt{DroidDetect\textsubscript{CLS}-Base}, trained with explicit exposure to such samples, maintains strong performance with a recall above 0.9. This shows that training on diverse, adversarially crafted examples further enhances robustness and reduces susceptibility to trivial evasion strategies.

\begin{table*}[!htbp]
\centering
\scalebox{0.55}{
\begin{tabular}{ll*{14}{r}}
\toprule
 & \textbf{\texttt{Model}} 
 & \multicolumn{7}{c}{\textbf{\texttt{2-Class}}} 
 & \multicolumn{7}{c}{\textbf{\texttt{3-Class}}} \\
\cmidrule(lr){3-9} \cmidrule(lr){10-16}
 &  
 & \texttt{\textbf{C/C++}} & \texttt{\textbf{C\#}} & \texttt{\textbf{Go}} & \texttt{\textbf{Java}} & \texttt{\textbf{Python}} & \texttt{\textbf{JS}} & \texttt{\textbf{Avg.}}
 & \texttt{\textbf{C/C++}} & \texttt{\textbf{C\#}} & \texttt{\textbf{Go}} & \texttt{\textbf{Java}} & \texttt{\textbf{Python}} & \texttt{\textbf{JS}} & \texttt{\textbf{Avg.}} \\
\midrule

\multirow{5}{*}{\texttt{Zero-Shot Baselines}} 
& \texttt{Fast-DetectGPT}~\citep{fastdetect}    & \texttt{81.33} & \texttt{72.77} & \texttt{81.16} & \texttt{76.03} & \texttt{73.60} & \texttt{74.59} & \texttt{76.58} & \texttt{77.85} & \texttt{66.37} & \texttt{72.73} & \texttt{69.45} & \texttt{70.34} & \texttt{69.11} & \texttt{70.98} \\
& \texttt{CoDet-M4}~\citep{orel2025codetm4detectingmachinegeneratedcode}          & \texttt{61.12} & \texttt{50.68} & \texttt{19.66} & \texttt{56.15} & \texttt{58.75} & \texttt{41.44} & \texttt{47.97} & \texttt{53.81} & \texttt{40.74} & \texttt{18.28} & \texttt{45.26} & \texttt{53.51} & \texttt{36.09} & \texttt{41.28} \\
&\texttt{M4}~\citep{m4}                 & \texttt{62.22} & \texttt{40.73} & \texttt{57.59} & \texttt{48.39} & \texttt{61.47} & \texttt{64.44} & \texttt{52.81} & \texttt{65.33} & \texttt{50.38} & \texttt{60.49} & \texttt{56.25} & \texttt{61.21} & \texttt{53.64} & \texttt{57.92} \\
& \texttt{GPTSniffer}~\citep{NGUYEN2024112059}         & \texttt{63.02} & \texttt{48.90} & \texttt{79.89} & \texttt{40.30} & \texttt{38.34} & \texttt{45.94} & \texttt{52.40} & \texttt{64.18} & \texttt{42.29} & \texttt{76.19} & \texttt{34.94} & \texttt{34.94} & \texttt{47.22} & \texttt{49.96} \\
& \texttt{GPTZero}            & \texttt{58.32} & \texttt{45.69} & \texttt{13.64} & \texttt{74.65} & \texttt{73.19} & \texttt{63.16} & \texttt{54.81} & \texttt{61.00} & \texttt{50.38} & \texttt{28.89} & \texttt{61.25} & \texttt{52.63} & \texttt{54.78} & \texttt{51.48} \\
\midrule

\multirow{6}{*}{\texttt{OOD Evaluations}}
& \texttt{DroidDetect\textsubscript{CLS}-Base\textsubscript{C/C++}}     & \texttt{98.98} & \texttt{96.59} & \texttt{67.32} & \texttt{96.97} & \texttt{74.45} & \texttt{91.15} & \texttt{87.58} & \texttt{92.62} & \texttt{81.67} & \texttt{56.43} & \texttt{79.45} & \texttt{56.43} & \texttt{69.72} & \texttt{72.72} \\
& \texttt{DroidDetect\textsubscript{CLS}-Base\textsubscript{C\#}}       & \texttt{93.66} & \texttt{99.20} & \texttt{78.89} & \texttt{95.20} & \texttt{71.13} & \texttt{89.87} & \texttt{87.99} & \texttt{80.95} & \texttt{92.93} & \texttt{57.74} & \texttt{84.17} & \texttt{54.25} & \texttt{65.18} & \texttt{71.04} \\
& \texttt{DroidDetect\textsubscript{CLS}-Base\textsubscript{Go}}        & \texttt{93.33} & \texttt{86.00} & \texttt{98.94} & \texttt{89.97} & \texttt{71.45} & \texttt{88.72} & \texttt{88.07} & \texttt{80.74} & \texttt{63.61} & \texttt{92.93} & \texttt{74.18} & \texttt{50.38} & \texttt{65.37} & \texttt{71.20} \\
& \texttt{DroidDetect\textsubscript{CLS}-Base\textsubscript{Java}}      & \texttt{95.53} & \texttt{96.42} & \texttt{94.57} & \texttt{99.31} & \texttt{75.59} & \texttt{80.26} & \texttt{90.28} & \texttt{85.00} & \texttt{84.43} & \texttt{58.85} & \texttt{93.38} & \texttt{63.25} & \texttt{64.57} & \texttt{74.91} \\
& \texttt{DroidDetect\textsubscript{CLS}-Base\textsubscript{Python}}    & \texttt{80.27} & \texttt{85.48} & \texttt{82.28} & \texttt{88.80} & \texttt{98.85} & \texttt{86.62} & \texttt{86.75} & \texttt{67.59} & \texttt{75.56} & \texttt{53.70} & \texttt{79.31} & \texttt{93.08} & \texttt{69.96} & \texttt{73.20} \\
& \texttt{DroidDetect\textsubscript{CLS}-Base\textsubscript{JS}}        & \texttt{95.76} & \texttt{97.38} & \texttt{75.27} & \texttt{96.45} & \texttt{68.98} & \texttt{97.80} & \texttt{88.61} & \texttt{87.96} & \texttt{87.58} & \texttt{52.78} & \texttt{86.32} & \texttt{60.78} & \texttt{89.67} & \texttt{77.52} \\
\midrule

\multirow{2}{*}{\texttt{Fine-Tuned Baselines}}
& \texttt{GCN}                & \texttt{79.06} & \texttt{78.33} & \texttt{84.33} & \texttt{80.04} & \texttt{72.49} & \texttt{69.69} & \texttt{77.32} & \texttt{65.97} & \texttt{58.03} & \texttt{65.20} & \texttt{60.13} & \texttt{55.22} & \texttt{54.72} & \texttt{59.88} \\
& \texttt{CatBoost}           & \texttt{94.00} & \texttt{91.20} & \texttt{90.57} & \texttt{92.26} & \texttt{89.51} & \texttt{82.55} & \texttt{90.02} & \texttt{84.57} & \texttt{81.32} & \texttt{81.54} & \texttt{82.42} & \texttt{78.15} & \texttt{70.98} & \texttt{78.83} \\
\midrule

\multirow{5}{*}{\texttt{Full Training}}
& \texttt{M4\textsubscript{FT}} & \texttt{94.18} & \texttt{89.98} & \texttt{92.19} & \texttt{92.63} & \texttt{87.19} & \texttt{93.61} & \texttt{91.63} & \texttt{79.56} & \texttt{75.55} & \texttt{77.63} & \texttt{79.55} & \texttt{69.63} & \texttt{77.53} & \texttt{76.57}\\

& \texttt{GPT-Sniffer\textsubscript{FT}} & \texttt{97.64} & \texttt{97.36} & \texttt{97.33} & \texttt{97.96} & \texttt{95.07} & \texttt{97.94} & \texttt{97.22} & \texttt{85.14} & \texttt{85.75} & \texttt{85.78} & \texttt{86.97} & \texttt{79.02} & \texttt{88.28} & \texttt{85.16} \\

& \texttt{CoDet-M4\textsubscript{FT}} & \texttt{99.36} & \texttt{99.22} & \texttt{99.31} & \texttt{99.04} & \texttt{98.28} & \texttt{99.24} & \texttt{99.08} & \texttt{89.98} & \texttt{88.94} & \texttt{89.73} & \texttt{91.70} & \texttt{85.80} & \texttt{91.46} & \texttt{89.60}\\

& \texttt{DroidDetect\textsubscript{CLS}-Base}          & \texttt{99.29} & \texttt{99.33} & \texttt{99.32} & \texttt{99.45} & \texttt{98.87} & \texttt{98.38} & \texttt{99.11} & \texttt{94.43} & \texttt{94.06} & \texttt{93.98} & \texttt{93.93} & \texttt{93.95} & \texttt{90.99} & \texttt{93.56} \\

& \texttt{DroidDetect\textsubscript{CLS}-Large}          & \texttt{\textbf{99.31}} & \texttt{\textbf{99.51}} & \texttt{\textbf{99.32}} & \texttt{\textbf{99.45}} & \texttt{\textbf{99.11}} & \texttt{\textbf{98.67}} & \texttt{\textbf{99.23}} & \texttt{\textbf{94.24}} & \texttt{\textbf{93.87}} & \texttt{\textbf{94.42}} & \texttt{\textbf{94.05}} & \texttt{\textbf{94.13}} & \texttt{\textbf{91.27}} & \texttt{\textbf{93.66}} \\
\bottomrule
\end{tabular}
}
\caption{Comparison of models in 2-class (human- vs. machine-generated) vs. 3-class (human- vs. machine-generated vs. machine-refined) classification setups across programming languages in terms of weighted F1-score. In the OOD section, we train on each programming language individually.
The best results are highlighted in \textbf{bold}.}
\label{tab:lang_eval}
\end{table*}

\begin{table*}[h]
\centering
\scalebox{0.5}{
\begin{tabular}{l| *{10}{c}}
\toprule
 & \texttt{\textbf{FastDetectGPT}} & \texttt{\textbf{GPTSniffer}} & \texttt{\textbf{M4}} & \texttt{\textbf{CoDet-M4}} & \texttt{\textbf{GPT-Zero}} & \texttt{\textbf{M4\textsubscript{FT}}} & \texttt{\textbf{GPT-Sniffer\textsubscript{FT}}} & \texttt{\textbf{CoDet-M4\textsubscript{FT}}}  & \texttt{\textbf{DroidDetect\textsubscript{CLS}-Base}} & \texttt{\textbf{DroidDetect\textsubscript{CLS}-Large}} \\
\midrule
\texttt{Human-written} & \texttt{0.84} & \texttt{0.65} & \cellcolor{red!20}\texttt{0.40} & \cellcolor{red!20}\texttt{0.38} & \texttt{0.53}  & \texttt{0.91} & \texttt{0.97} & \texttt{0.96} &  \texttt{0.93} & \texttt{\textbf{0.98}}\\
\texttt{Adversarial samples} & \texttt{0.48} & \texttt{0.49} & \texttt{0.73} & \texttt{0.63} & \texttt{0.10} & \texttt{0.67} & \texttt{0.55} & \texttt{0.51} & \texttt{\textbf{0.92}} &  \texttt{\textbf{0.92}}\\
\bottomrule
\end{tabular}
}
\caption{Recall for human-written vs. adversarial examples. The red cells show that despite having high recall on adversarial samples, M4 and CoDet-M4 struggle to detect human-written code. The best results are in \textbf{bold}.}
\label{tab:adversarial_eval}
\end{table*}

\section{Detector Training and Ablations}
\label{sec:training}

We conducted a series of ablation experiments starting from our \texttt{DroidDetect\textsubscript{CLS}} backbone to systematically identify the most effective model architecture and training strategy.

\blue{As an architectural ablation,} we explore whether incorporating the structural representation of code could improve the detector's performance. 
Specifically, we trained a 4-layer Graph Convolutional Network over the AST representation of codes to evaluate its ability to distinguish AI-generated from human-written code.
The results are shown in \Cref{appx:gnn}. We can see that while structural signals are informative, GCNs alone are not sufficient to achieve strong generalisation.

Next, we explored early fusion of textual and structural representations by combining a text encoder with a GCN encoder. 
For text encoding, we used ModernBERT~\citep{modernbert}, a transformer-based model pre-trained on a mixture of natural language and code. 
We experimented with both the base (149M parameters) and the large (396M parameters) variants. 
This model was selected for the inference efficiency~\citep{modernbert}, and suitability for code-related tasks.
However, as shown in \Cref{appx:fusion_vs_no_fusion}, this fusion strategy yielded only a marginal improvement. 
Consequently, we decided to use a text-only encoder for the final model.

We then address the issue of class separability, which can arise because adversarial and refined code \blue{is} similar to human-written code. We explore training our models using triplet loss~\citep{tripletloss} in a supervised contrastive~\citep{DBLP:conf/nips/KhoslaTWSTIMLK20} setup using the class labels. This metric-learning approach encourages the model to place samples of the same class closer to each other in the embedding space while pushing dissimilar samples apart, and it has been demonstrably effective in other detection scenarios that require high precision~\citep{DBLP:journals/pami/DengGYXKZ22,DBLP:conf/acl/LiL24}. We refer to these models as \texttt{DroidDetect\textsubscript{SCL}}. \Cref{tab:base_vs_large_models} demonstrates the small but consistent performance gain unlocked using metric learning.

Finally, we addressed the problem of noisy and mislabelled training data. Despite extensive data filtering, it is possible that some code samples curated as human-written may have been generated by coding-copilots. The presence of such examples could negatively impact the training of our detector. To address this, we applied MC Dropout~\citep{uncertainty} to estimate the model uncertainty on the human-written portion of the test set.
Specifically, we identified the top 7\% most uncertain samples-- those for which a pre-trained model exhibited low prediction confidence--and resampled the dataset, removing them from the training set. We then retrained the model on the remaining data, thereby getting rid of the influence of potentially mislabelled or ambiguous samples. This manner of self-bootstrapping datasets has an extensive track record in image~\citep{DBLP:journals/corr/abs-1905-00546,DBLP:conf/cvpr/XieLHL20} and text~\citep{DBLP:journals/corr/abs-2212-03533} representation learning, relying on the tendency of neural networks to understand patterns in clean labels before they overfit to noisy data~\citep{DBLP:conf/nips/FeldmanZ20}. Incorporating this filtering into our training yielded our final \texttt{DroidDetect} models, which, as \blue{shown} in \Cref{tab:base_vs_large_models}, perform the best across the classification settings. 

We trained all models for 3 epochs, using AdamW~\citep{adamw} optimiser, setting the top learning rate to 5e-5, and applying the linear warmup (proportion 0.1) with cosine decay learning rate scheduler. The batch size is 64 for \texttt{DroidDetect-Base} and 40 for \texttt{DroidDetect-Large}.

\begin{table}[h]
\centering
\scalebox{0.65}{
\begin{tabular}{lcccccc}
\toprule
 \texttt{\textbf{Model Variant}} & \multicolumn{2}{c}{\texttt{\textbf{2-class}}} & \multicolumn{2}{c}{\texttt{\textbf{3-class}}} & \multicolumn{2}{c}{\texttt{\textbf{4-class}}} \\
\cmidrule(lr){2-3} \cmidrule(lr){4-5} \cmidrule(lr){6-7}
 & \texttt{\textbf{Base}} & \texttt{\textbf{Large}} & \texttt{\textbf{Base}} & \texttt{\textbf{Large}} & \texttt{\textbf{Base}} & \texttt{\textbf{Large}} \\
\midrule
\texttt{DroidDetect} &  \texttt{\textbf{99.18}} &  \texttt{\textbf{99.25}} & \texttt{\textbf{94.36}} & \texttt{\textbf{95.17}} & \texttt{\textbf{92.95}} & \texttt{\textbf{94.30}} \\
\midrule
\texttt{- Resampling} & \multirow{2}{*}{\texttt{99.15}} & \multirow{2}{*}{\texttt{99.22}} & \multirow{2}{*}{\texttt{93.86}} & \multirow{2}{*}{\texttt{94.43}} & \multirow{2}{*}{\texttt{92.52}} & \multirow{2}{*}{\texttt{93.14}} \\
\small{\hspace{1.2em}[\texttt{DroidDetect\textsubscript{SCL}}]} & & & & & &  \\
\midrule
\texttt{- Triplet Loss} & \multirow{2}{*}{\texttt{99.14}} & \multirow{2}{*}{\texttt{99.18}} & \multirow{2}{*}{\texttt{90.51}} & \multirow{2}{*}{\texttt{94.07}} & \multirow{2}{*}{\texttt{89.63}} & \multirow{2}{*}{\texttt{92.65}} \\
\small{\hspace{1.2em}[\texttt{DroidDetect\textsubscript{CLS}}]} & & & & & &  \\
\bottomrule
\end{tabular}}
\caption{Weighted F1-score for \texttt{DroidDetect} across training ablations. The best results are shown in \textbf{bold}.}
\label{tab:base_vs_large_models}
\end{table}

\section{Conclusion and Future Work}
\label{sec:Conclusion}

We have curated \texttt{DroidCollection}, a large and diverse suite of datasets that facilitate the training and the evaluation of robust AI-generated code detectors to support their most common modes of operation, i.e., completion and rewriting, along with potentially adversarial use cases. 
Among openly available corpora for training AI-generated content detectors, \texttt{DroidCollection} offers the most exhaustive coverage with respect to the number of generators, generation settings, programming languages, and domains covered. We further developed \texttt{DroidDetect}, a suite of AI-assisted code detection models, in two sizes (Base and Large), and conducted extensive ablation studies to evaluate which training strategies yield the most effective performance for this task.

In future work, we plan to enhance the coverage of \texttt{DroidCollection} and the robustness of \texttt{DroidDetect}. We will incorporate code samples from more closed-source API-based generators, broadening the diversity of the code samples. We will also add generations from reasoning LMs to enhance the applicability of our detectors. Finally, we plan to expand language coverage to include languages such as PHP, Rust, and Ruby, thereby making our benchmarks more representative.


\section*{Limitations}
\paragraph{Corpus Updates and Coverage} Possessing a perfect coverage over all major models in the current fast-paced release environment is an intractable task. We acknowledge that the release of new model families with unseen output distributions presents a challenge for all AI-generated content detectors. Since we have mature pipelines for machine-generated, machine-refined and adversarially-humanised data acquisition, we plan to update \texttt{DroidCollection} with generations sourced from future model releases. 

\paragraph{Cost Effectiveness} Owing to cost realities, the majority of training samples in our study are sourced from locally deployable models. The high costs of API invocations are the primary reason why our study leaves data collection from recently released reasoning/thinking models such as Anthropic’s \texttt{Claude~3.7}, DeepSeek~\texttt{R1}, and Google’s \texttt{Gemini~2.5} for future work. For similar reasons, our evaluation of API-based detectors such as GPTZero was limited to a subset of the test set. 

\paragraph{Potential Data Contamination} In spite of the thorough curation and extensive filtering undertaken for \texttt{DroidCollection}, we acknowledge the possibility that a small number of AI-generated or AI-assisted code samples may still be mislabeled as human-authored, due to the inherent nature of the data sources used for the dataset construction. 
Seeking to limit the negative effects of mislabeled or noisy data, our work explores uncertainty-based dataset re-sampling using a pre-trained classifier, which we show to be effective in improving the model's performance by identifying ambiguous samples to discard during training. 
In the released dataset, we include flags for code snippets identified as suspicious, enabling downstream users to apply additional filtering or analysis as needed. 

\section*{Ethics Statement}
The human-written code samples in our dataset are sourced exclusively from publicly available code corpora vetted for appropriate licensing and PII removal. Additionally, all code generation was conducted in compliance with the terms of use of the respective model providers.

\texttt{DroidDetect} and \texttt{DroidCollection} aim to promote transparency in code authorship, especially in academic and research settings. While there is a risk that they could be misused to train models to evade detection, we strongly discourage any malicious or privacy-invasive applications. We advocate for the responsible use in strictly legitimate research and educational contexts.

\section*{Acknowledgements}

At TU Darmstadt, this research work has been supported by the German Federal Ministry of Research, Technology and Space and the Hessian Ministry of Higher Education, Research, Science and the Arts within their joint support of the National Research Center for Applied Cybersecurity ATHENE and by HUAWEI Technologies (Ireland) Co., Ltd.

\bibliography{acl_latex}

\begin{thebibliography}{90}
\providecommand{\natexlab}[1]{#1}

\bibitem[{Abassy et~al.(2024)Abassy, Elozeiri, Aziz, Ta, Tomar, Adhikari, Ahmed, Wang, Afzal, Xie, Mansurov, Artemova, Mikhailov, Xing, Geng, Iqbal, Mujahid, Mahmoud, Tsvigun, Aji, Shelmanov, Habash, Gurevych, and Nakov}]{DBLP:journals/corr/abs-2408-04284}
Mervat Abassy, Kareem~Ashraf Elozeiri, Alexander Aziz, Minh~Ngoc Ta, Raj~Vardhan Tomar, Bimarsha Adhikari, Saad El~Dine Ahmed, Yuxia Wang, Osama~Mohammed Afzal, Zhuohan Xie, Jonibek Mansurov, Ekaterina Artemova, Vladislav Mikhailov, Rui Xing, Jiahui Geng, Hasan Iqbal, Zain~Muhammad Mujahid, Tarek Mahmoud, Akim Tsvigun, and 5 others. 2024.
\newblock \href {https://doi.org/10.48550/ARXIV.2408.04284} {{LLM-DetectAIve}: a tool for fine-grained machine-generated text detection}.
\newblock \emph{CoRR}, abs/2408.04284.

\bibitem[{Abdin et~al.(2024)Abdin, Aneja, Behl, Bubeck, Eldan, Gunasekar, Harrison, Hewett, Javaheripi, Kauffmann, Lee, Lee, Li, Liu, Mendes, Nguyen, Price, de~Rosa, Saarikivi, Salim, Shah, Wang, Ward, Wu, Yu, Zhang, and Zhang}]{abdin2024phi4technicalreport}
Marah Abdin, Jyoti Aneja, Harkirat Behl, Sébastien Bubeck, Ronen Eldan, Suriya Gunasekar, Michael Harrison, Russell~J. Hewett, Mojan Javaheripi, Piero Kauffmann, James~R. Lee, Yin~Tat Lee, Yuanzhi Li, Weishung Liu, Caio C.~T. Mendes, Anh Nguyen, Eric Price, Gustavo de~Rosa, Olli Saarikivi, and 8 others. 2024.
\newblock \href {https://arxiv.org/abs/2412.08905} {{Phi-4 Technical Report}}.
\newblock \emph{Preprint}, arXiv:2412.08905.

\bibitem[{AI et~al.(2025)AI, :, Young, Chen, Li, Huang, Zhang, Zhang, Wang, Li, Zhu, Chen, Chang, Yu, Liu, Liu, Yue, Yang, Yang, Xie, Huang, Hu, Ren, Niu, Nie, Li, Xu, Liu, Wang, Cai, Gu, Liu, and Dai}]{Young2024YiOF}
01. AI, :, Alex Young, Bei Chen, Chao Li, Chengen Huang, Ge~Zhang, Guanwei Zhang, Guoyin Wang, Heng Li, Jiangcheng Zhu, Jianqun Chen, Jing Chang, Kaidong Yu, Peng Liu, Qiang Liu, Shawn Yue, Senbin Yang, Shiming Yang, and 14 others. 2025.
\newblock \href {https://arxiv.org/abs/2403.04652} {{Yi: Open Foundation Models by 01.AI}}.
\newblock \emph{Preprint}, arXiv:2403.04652.

\bibitem[{AI(2025)}]{mistral2025}
Mistral AI. 2025.
\newblock {Mistral Small – A new balance of performance and efficiency}.
\newblock Online.
\newblock Available at \url{https://mistral.ai/news/mistral-small-3} (Accessed: 1 April 2025).

\bibitem[{Ancona et~al.(2018)Ancona, Ceolini, {\"{O}}ztireli, and Gross}]{grad_attrib}
Marco Ancona, Enea Ceolini, Cengiz {\"{O}}ztireli, and Markus Gross. 2018.
\newblock \href {https://openreview.net/forum?id=Sy21R9JAW} {Towards better understanding of gradient-based attribution methods for deep neural networks}.
\newblock In \emph{6th International Conference on Learning Representations, {ICLR} 2018, Vancouver, BC, Canada, April 30 - May 3, 2018, Conference Track Proceedings}. OpenReview.net.

\bibitem[{Bao et~al.(2024)Bao, Zhao, Teng, Yang, and Zhang}]{fastdetect}
Guangsheng Bao, Yanbin Zhao, Zhiyang Teng, Linyi Yang, and Yue Zhang. 2024.
\newblock \href {https://openreview.net/forum?id=Bpcgcr8E8Z} {{Fast-DetectGPT: Efficient Zero-Shot Detection of Machine-Generated Text via Conditional Probability Curvature}}.
\newblock In \emph{The Twelfth International Conference on Learning Representations, {ICLR} 2024, Vienna, Austria, May 7-11, 2024}. OpenReview.net.

\bibitem[{Bavarian et~al.(2022)Bavarian, Jun, Tezak, Schulman, McLeavey, Tworek, and Chen}]{DBLP:journals/corr/abs-2207-14255}
Mohammad Bavarian, Heewoo Jun, Nikolas Tezak, John Schulman, Christine McLeavey, Jerry Tworek, and Mark Chen. 2022.
\newblock \href {https://doi.org/10.48550/ARXIV.2207.14255} {{Efficient Training of Language Models to Fill in the Middle}}.
\newblock \emph{CoRR}, abs/2207.14255.

\bibitem[{Broder(1997)}]{minhash}
A.Z. Broder. 1997.
\newblock \href {https://doi.org/10.1109/SEQUEN.1997.666900} {On the resemblance and containment of documents}.
\newblock In \emph{Proceedings. Compression and Complexity of SEQUENCES 1997 (Cat. No.97TB100171)}, pages 21--29.

\bibitem[{Bukhari(2024)}]{bukhari2024issues}
Sufiyan~Ahmed Bukhari. 2024.
\newblock {Issues in Detection of AI-Generated Source Code}.
\newblock \emph{University of Calgary}.

\bibitem[{Chamezopoulos et~al.(2024)Chamezopoulos, Herrmannova, De~Waard, Herrmannova, Rosati, and Kashnitsky}]{dagap}
Savvas Chamezopoulos, Drahomira Herrmannova, Anita De~Waard, Drahomira Herrmannova, Domenic Rosati, and Yury Kashnitsky. 2024.
\newblock \href {https://aclanthology.org/2024.sdp-1.2/} {{Overview of the {D}ag{P}ap24 Shared Task on Detecting Automatically Generated Scientific Paper}}.
\newblock In \emph{Proceedings of the Fourth Workshop on Scholarly Document Processing (SDP 2024)}, pages 7--11, Bangkok, Thailand. Association for Computational Linguistics.

\bibitem[{Clark et~al.(2019)Clark, Khandelwal, Levy, and Manning}]{DBLP:conf/blackboxnlp/ClarkKLM19}
Kevin Clark, Urvashi Khandelwal, Omer Levy, and Christopher~D. Manning. 2019.
\newblock \href {https://doi.org/10.18653/V1/W19-4828} {What does {BERT} look at? an analysis of bert's attention}.
\newblock In \emph{Proceedings of the 2019 {ACL} Workshop BlackboxNLP: Analyzing and Interpreting Neural Networks for NLP, BlackboxNLP@ACL 2019, Florence, Italy, August 1, 2019}, pages 276--286. Association for Computational Linguistics.

\bibitem[{Deng et~al.(2022)Deng, Guo, Yang, Xue, Kotsia, and Zafeiriou}]{DBLP:journals/pami/DengGYXKZ22}
Jiankang Deng, Jia Guo, Jing Yang, Niannan Xue, Irene Kotsia, and Stefanos Zafeiriou. 2022.
\newblock \href {https://doi.org/10.1109/TPAMI.2021.3087709} {{ArcFace: Additive Angular Margin Loss for Deep Face Recognition}}.
\newblock \emph{{IEEE} Trans. Pattern Anal. Mach. Intell.}, 44(10):5962--5979.

\bibitem[{Dugan et~al.(2024)Dugan, Hwang, Trhl{\'i}k, Zhu, Ludan, Xu, Ippolito, and Callison-Burch}]{raid}
Liam Dugan, Alyssa Hwang, Filip Trhl{\'i}k, Andrew Zhu, Josh~Magnus Ludan, Hainiu Xu, Daphne Ippolito, and Chris Callison-Burch. 2024.
\newblock \href {https://doi.org/10.18653/v1/2024.acl-long.674} {{{RAID}: A Shared Benchmark for Robust Evaluation of Machine-Generated Text Detectors}}.
\newblock In \emph{Proceedings of the 62nd Annual Meeting of the Association for Computational Linguistics (Volume 1: Long Papers)}, pages 12463--12492, Bangkok, Thailand. Association for Computational Linguistics.

\bibitem[{Dunay et~al.(2024)Dunay, Cheng, Tait, Thakkar, Rigby, Chiu, Ahmad, Ganesan, Maddila, Murali, Tayyebi, and Nagappan}]{DBLP:conf/sigsoft/DunayCTTRCAGMMT24}
Omer Dunay, Daniel Cheng, Adam Tait, Parth Thakkar, Peter~C. Rigby, Andy Chiu, Imad Ahmad, Arun Ganesan, Chandra~Shekhar Maddila, Vijayaraghavan Murali, Ali Tayyebi, and Nachiappan Nagappan. 2024.
\newblock \href {https://doi.org/10.1145/3663529.3663836} {{Multi-line AI-Assisted Code Authoring}}.
\newblock In \emph{Companion Proceedings of the 32nd {ACM} International Conference on the Foundations of Software Engineering, {FSE} 2024, Porto de Galinhas, Brazil, July 15-19, 2024}, pages 150--160. {ACM}.

\bibitem[{Feldman and Zhang(2020)}]{DBLP:conf/nips/FeldmanZ20}
Vitaly Feldman and Chiyuan Zhang. 2020.
\newblock \href {https://proceedings.neurips.cc/paper/2020/hash/1e14bfe2714193e7af5abc64ecbd6b46-Abstract.html} {{What Neural Networks Memorize and Why: Discovering the Long Tail via Influence Estimation}}.
\newblock In \emph{Advances in Neural Information Processing Systems 33: Annual Conference on Neural Information Processing Systems 2020, NeurIPS 2020, December 6-12, 2020, virtual}.

\bibitem[{Fr{\"{o}}mmgen et~al.(2024)Fr{\"{o}}mmgen, Austin, Choy, Ghelani, Kharatyan, Surita, Khrapko, Lamblin, Manzagol, Revaj, Tabachnyk, Tarlow, Villela, Zheng, Chandra, and Maniatis}]{DBLP:conf/icse/FrommgenA24}
Alexander Fr{\"{o}}mmgen, Jacob Austin, Peter Choy, Nimesh Ghelani, Lera Kharatyan, Gabriela Surita, Elena Khrapko, Pascal Lamblin, Pierre{-}Antoine Manzagol, Marcus Revaj, Maxim Tabachnyk, Daniel Tarlow, Kevin Villela, Daniel Zheng, Satish Chandra, and Petros Maniatis. 2024.
\newblock \href {https://doi.org/10.1145/3639477.3639746} {{Resolving Code Review Comments with Machine Learning}}.
\newblock In \emph{Proceedings of the 46th International Conference on Software Engineering: Software Engineering in Practice, {ICSE-SEIP} 2024, Lisbon, Portugal, April 14-20, 2024}, pages 204--215. {ACM}.

\bibitem[{Fujii et~al.(2025)Fujii, Tajima, Mizuki, Shimada, Shiotani, Saito, Ohi, Kawamura, Nakamura, Okamoto, Ishida, Hattori, Ma, Takamura, Yokota, and Okazaki}]{swallowcode}
Kazuki Fujii, Yukito Tajima, Sakae Mizuki, Hinari Shimada, Taihei Shiotani, Koshiro Saito, Masanari Ohi, Masaki Kawamura, Taishi Nakamura, Takumi Okamoto, Shigeki Ishida, Kakeru Hattori, Youmi Ma, Hiroya Takamura, Rio Yokota, and Naoaki Okazaki. 2025.
\newblock \href {https://arxiv.org/abs/2505.02881} {{Rewriting Pre-Training Data Boosts LLM Performance in Math and Code}}.
\newblock \emph{Preprint}, arXiv:2505.02881.

\bibitem[{Ge et~al.(2024)Ge, Chan, Wang, Yu, Mi, and Yu}]{persona}
Tao Ge, Xin Chan, Xiaoyang Wang, Dian Yu, Haitao Mi, and Dong Yu. 2024.
\newblock \href {https://arxiv.org/abs/2406.20094} {{Scaling Synthetic Data Creation with 1,000,000,000 Personas}}.
\newblock \emph{Preprint}, arXiv:2406.20094.

\bibitem[{Gemma et~al.(2024)Gemma, Mesnard, Hardin, Dadashi, Bhupatiraju, Pathak, Sifre, Rivière, Kale, Love, Tafti, Hussenot, Sessa, Chowdhery, Roberts, Barua, Botev, Castro-Ros, Slone, Héliou, Tacchetti, Bulanova, Paterson, Tsai, Shahriari, Lan, Choquette-Choo, Crepy, Cer, Ippolito, Reid, Buchatskaya, Ni, Noland, Yan, Tucker, Muraru, Rozhdestvenskiy, Michalewski, Tenney, Grishchenko, Austin, Keeling, Labanowski, Lespiau, Stanway, Brennan, Chen, Ferret, Chiu, Mao-Jones, Lee, Yu, Millican, Sjoesund, Lee, Dixon, Reid, Mikuła, Wirth, Sharman, Chinaev, Thain, Bachem, Chang, Wahltinez, Bailey, Michel, Yotov, Chaabouni, Comanescu, Jana, Anil, McIlroy, Liu, Mullins, Smith, Borgeaud, Girgin, Douglas, Pandya, Shakeri, De, Klimenko, Hennigan, Feinberg, Stokowiec, hui Chen, Ahmed, Gong, Warkentin, Peran, Giang, Farabet, Vinyals, Dean, Kavukcuoglu, Hassabis, Ghahramani, Eck, Barral, Pereira, Collins, Joulin, Fiedel, Senter, Andreev, and Kenealy}]{team2024gemma}
Team Gemma, Thomas Mesnard, Cassidy Hardin, Robert Dadashi, Surya Bhupatiraju, Shreya Pathak, Laurent Sifre, Morgane Rivière, Mihir~Sanjay Kale, Juliette Love, Pouya Tafti, Léonard Hussenot, Pier~Giuseppe Sessa, Aakanksha Chowdhery, Adam Roberts, Aditya Barua, Alex Botev, Alex Castro-Ros, Ambrose Slone, and 89 others. 2024.
\newblock \href {https://arxiv.org/abs/2403.08295} {{Gemma: Open Models Based on Gemini Research and Technology}}.
\newblock \emph{Preprint}, arXiv:2403.08295.

\bibitem[{Grattafiori et~al.(2024)Grattafiori, Dubey, Jauhri, Pandey, Kadian, Al-Dahle, Letman, Mathur, Schelten, Vaughan, Yang, Fan, Goyal, Hartshorn, Yang, Mitra, Sravankumar, Korenev, Hinsvark, Rao, Zhang, Rodriguez, Gregerson, Spataru, Roziere, Biron, Tang, Chern, Caucheteux, Nayak, Bi, Marra, McConnell, Keller, Touret, Wu, Wong, Ferrer, Nikolaidis, Allonsius, Song, Pintz, Livshits, Wyatt, Esiobu, Choudhary, Mahajan, Garcia-Olano, Perino, Hupkes, Lakomkin, AlBadawy, Lobanova, Dinan, Smith, Radenovic, Guzmán, Zhang, Synnaeve, Lee, Anderson, Thattai, Nail, Mialon, Pang, Cucurell, Nguyen, Korevaar, Xu, Touvron, Zarov, Ibarra, Kloumann, Misra, Evtimov, Zhang, Copet, Lee, Geffert, Vranes, Park, Mahadeokar, Shah, van~der Linde, Billock, Hong, Lee, Fu, Chi, Huang, Liu, Wang, Yu, Bitton, Spisak, Park, Rocca, Johnstun, Saxe, Jia, Alwala, Prasad, Upasani, Plawiak, Li, Heafield, Stone, El-Arini, Iyer, Malik, Chiu, Bhalla, Lakhotia, Rantala-Yeary, van~der Maaten, Chen, Tan, Jenkins, Martin, Madaan, Malo, Blecher,
  Landzaat, de~Oliveira, Muzzi, Pasupuleti, Singh, Paluri, Kardas, Tsimpoukelli, Oldham, Rita, Pavlova, Kambadur, Lewis, Si, Singh, Hassan, Goyal, Torabi, Bashlykov, Bogoychev, Chatterji, Zhang, Duchenne, Çelebi, Alrassy, Zhang, Li, Vasic, Weng, Bhargava, Dubal, Krishnan, Koura, Xu, He, Dong, Srinivasan, Ganapathy, Calderer, Cabral, Stojnic, Raileanu, Maheswari, Girdhar, Patel, Sauvestre, Polidoro, Sumbaly, Taylor, Silva, Hou, Wang, Hosseini, Chennabasappa, Singh, Bell, Kim, Edunov, Nie, Narang, Raparthy, Shen, Wan, Bhosale, Zhang, Vandenhende, Batra, Whitman, Sootla, Collot, Gururangan, Borodinsky, Herman, Fowler, Sheasha, Georgiou, Scialom, Speckbacher, Mihaylov, Xiao, Karn, Goswami, Gupta, Ramanathan, Kerkez, Gonguet, Do, Vogeti, Albiero, Petrovic, Chu, Xiong, Fu, Meers, Martinet, Wang, Wang, Tan, Xia, Xie, Jia, Wang, Goldschlag, Gaur, Babaei, Wen, Song, Zhang, Li, Mao, Coudert, Yan, Chen, Papakipos, Singh, Srivastava, Jain, Kelsey, Shajnfeld, Gangidi, Victoria, Goldstand, Menon, Sharma, Boesenberg,
  Baevski, Feinstein, Kallet, Sangani, Teo, Yunus, Lupu, Alvarado, Caples, Gu, Ho, Poulton, Ryan, Ramchandani, Dong, Franco, Goyal, Saraf, Chowdhury, Gabriel, Bharambe, Eisenman, Yazdan, James, Maurer, Leonhardi, Huang, Loyd, Paola, Paranjape, Liu, Wu, Ni, Hancock, Wasti, Spence, Stojkovic, Gamido, Montalvo, Parker, Burton, Mejia, Liu, Wang, Kim, Zhou, Hu, Chu, Cai, Tindal, Feichtenhofer, Gao, Civin, Beaty, Kreymer, Li, Adkins, Xu, Testuggine, David, Parikh, Liskovich, Foss, Wang, Le, Holland, Dowling, Jamil, Montgomery, Presani, Hahn, Wood, Le, Brinkman, Arcaute, Dunbar, Smothers, Sun, Kreuk, Tian, Kokkinos, Ozgenel, Caggioni, Kanayet, Seide, Florez, Schwarz, Badeer, Swee, Halpern, Herman, Sizov, Guangyi, Zhang, Lakshminarayanan, Inan, Shojanazeri, Zou, Wang, Zha, Habeeb, Rudolph, Suk, Aspegren, Goldman, Zhan, Damlaj, Molybog, Tufanov, Leontiadis, Veliche, Gat, Weissman, Geboski, Kohli, Lam, Asher, Gaya, Marcus, Tang, Chan, Zhen, Reizenstein, Teboul, Zhong, Jin, Yang, Cummings, Carvill, Shepard, McPhie,
  Torres, Ginsburg, Wang, Wu, U, Saxena, Khandelwal, Zand, Matosich, Veeraraghavan, Michelena, Li, Jagadeesh, Huang, Chawla, Huang, Chen, Garg, A, Silva, Bell, Zhang, Guo, Yu, Moshkovich, Wehrstedt, Khabsa, Avalani, Bhatt, Mankus, Hasson, Lennie, Reso, Groshev, Naumov, Lathi, Keneally, Liu, Seltzer, Valko, Restrepo, Patel, Vyatskov, Samvelyan, Clark, Macey, Wang, Hermoso, Metanat, Rastegari, Bansal, Santhanam, Parks, White, Bawa, Singhal, Egebo, Usunier, Mehta, Laptev, Dong, Cheng, Chernoguz, Hart, Salpekar, Kalinli, Kent, Parekh, Saab, Balaji, Rittner, Bontrager, Roux, Dollar, Zvyagina, Ratanchandani, Yuvraj, Liang, Alao, Rodriguez, Ayub, Murthy, Nayani, Mitra, Parthasarathy, Li, Hogan, Battey, Wang, Howes, Rinott, Mehta, Siby, Bondu, Datta, Chugh, Hunt, Dhillon, Sidorov, Pan, Mahajan, Verma, Yamamoto, Ramaswamy, Lindsay, Lindsay, Feng, Lin, Zha, Patil, Shankar, Zhang, Zhang, Wang, Agarwal, Sajuyigbe, Chintala, Max, Chen, Kehoe, Satterfield, Govindaprasad, Gupta, Deng, Cho, Virk, Subramanian, Choudhury,
  Goldman, Remez, Glaser, Best, Koehler, Robinson, Li, Zhang, Matthews, Chou, Shaked, Vontimitta, Ajayi, Montanez, Mohan, Kumar, Mangla, Ionescu, Poenaru, Mihailescu, Ivanov, Li, Wang, Jiang, Bouaziz, Constable, Tang, Wu, Wang, Wu, Gao, Kleinman, Chen, Hu, Jia, Qi, Li, Zhang, Zhang, Adi, Nam, Yu, Wang, Zhao, Hao, Qian, Li, He, Rait, DeVito, Rosnbrick, Wen, Yang, Zhao, and Ma}]{llama3.1}
Aaron Grattafiori, Abhimanyu Dubey, Abhinav Jauhri, Abhinav Pandey, Abhishek Kadian, Ahmad Al-Dahle, Aiesha Letman, Akhil Mathur, Alan Schelten, Alex Vaughan, Amy Yang, Angela Fan, Anirudh Goyal, Anthony Hartshorn, Aobo Yang, Archi Mitra, Archie Sravankumar, Artem Korenev, Arthur Hinsvark, and 542 others. 2024.
\newblock \href {https://arxiv.org/abs/2407.21783} {{The {Llama} 3 Herd of Models}}.
\newblock \emph{Preprint}, arXiv:2407.21783.

\bibitem[{Guo et~al.(2023)Guo, Zhang, Wang, Jiang, Nie, Ding, Yue, and Wu}]{hc3}
Biyang Guo, Xin Zhang, Ziyuan Wang, Minqi Jiang, Jinran Nie, Yuxuan Ding, Jianwei Yue, and Yupeng Wu. 2023.
\newblock {How Close is ChatGPT to Human Experts? Comparison Corpus, Evaluation, and Detection}.
\newblock \emph{arXiv preprint arxiv:2301.07597}.

\bibitem[{Guo et~al.(2024)Guo, Zhu, Yang, Xie, Dong, Zhang, Chen, Bi, Wu, Li, Luo, Xiong, and Liang}]{guo2024deepseekcoderlargelanguagemodel}
Daya Guo, Qihao Zhu, Dejian Yang, Zhenda Xie, Kai Dong, Wentao Zhang, Guanting Chen, Xiao Bi, Y.~Wu, Y.~K. Li, Fuli Luo, Yingfei Xiong, and Wenfeng Liang. 2024.
\newblock \href {https://doi.org/10.48550/ARXIV.2401.14196} {{DeepSeek-Coder}: When the large language model meets programming - the rise of code intelligence}.
\newblock \emph{CoRR}, abs/2401.14196.

\bibitem[{Guo et~al.(2025)Guo, Cheng, Zhang, Shen, and Zhang}]{codemirage}
Hanxi Guo, Siyuan Cheng, Kaiyuan Zhang, Guangyu Shen, and Xiangyu Zhang. 2025.
\newblock \href {https://arxiv.org/abs/2506.11059} {{CodeMirage: A Multi-Lingual Benchmark for Detecting AI-Generated and Paraphrased Source Code from Production-Level LLMs}}.
\newblock \emph{Preprint}, arXiv:2506.11059.

\bibitem[{Hasan et~al.(2022)Hasan, Khosravi, Hossain, Rahman, and Nahavandi}]{uncertainty}
Md~Mehedi Hasan, Abbas Khosravi, Ibrahim Hossain, Ashikur Rahman, and Saeid Nahavandi. 2022.
\newblock \href {https://api.semanticscholar.org/CorpusID:248562494} {{Controlled Dropout for Uncertainty Estimation}}.
\newblock \emph{2023 IEEE International Conference on Systems, Man, and Cybernetics (SMC)}, pages 973--980.

\bibitem[{He et~al.(2024)He, Shen, Chen, Backes, and Zhang}]{DBLP:conf/ccs/0001SC0024}
Xinlei He, Xinyue Shen, Zeyuan Chen, Michael Backes, and Yang Zhang. 2024.
\newblock \href {https://doi.org/10.1145/3658644.3670344} {{{MGTBench}: Benchmarking Machine-Generated Text Detection}}.
\newblock In \emph{Proceedings of the 2024 on {ACM} {SIGSAC} Conference on Computer and Communications Security, {CCS} 2024, Salt Lake City, UT, USA, October 14-18, 2024}, pages 2251--2265. {ACM}.

\bibitem[{Hoffer and Ailon(2015)}]{tripletloss}
Elad Hoffer and Nir Ailon. 2015.
\newblock \href {http://arxiv.org/abs/1412.6622} {{Deep metric learning using Triplet network}}.
\newblock In \emph{3rd International Conference on Learning Representations, {ICLR} 2015, San Diego, CA, USA, May 7-9, 2015, Workshop Track Proceedings}.

\bibitem[{Hu et~al.(2022)Hu, Shen, Wallis, Allen-Zhu, Li, Wang, Wang, and Chen}]{hu2022lora}
Edward~J Hu, Yelong Shen, Phillip Wallis, Zeyuan Allen-Zhu, Yuanzhi Li, Shean Wang, Lu~Wang, and Weizhu Chen. 2022.
\newblock \href {https://openreview.net/forum?id=nZeVKeeFYf9} {{Lo{RA}: Low-Rank Adaptation of Large Language Models}}.
\newblock In \emph{International Conference on Learning Representations}.

\bibitem[{Husain et~al.(2020)Husain, Wu, Gazit, Allamanis, and Brockschmidt}]{codesearchnet}
Hamel Husain, Ho-Hsiang Wu, Tiferet Gazit, Miltiadis Allamanis, and Marc Brockschmidt. 2020.
\newblock \href {https://arxiv.org/abs/1909.09436} {{CodeSearchNet Challenge: Evaluating the State of Semantic Code Search}}.
\newblock \emph{Preprint}, arXiv:1909.09436.

\bibitem[{Idialu et~al.(2024)Idialu, Mathews, Maipradit, Atlee, and Nagappan}]{whodunit}
Oseremen~Joy Idialu, Noble~Saji Mathews, Rungroj Maipradit, Joanne~M. Atlee, and Mei Nagappan. 2024.
\newblock \href {https://doi.org/10.1145/3643991.3644926} {{Whodunit: Classifying Code as Human Authored or GPT-4 Generated - A case study on CodeChef problems}}.
\newblock In \emph{Proceedings of the 21st International Conference on Mining Software Repositories}, MSR '24, page 394–406, New York, NY, USA. Association for Computing Machinery.

\bibitem[{Ippolito et~al.(2020)Ippolito, Duckworth, Callison-Burch, and Eck}]{sampling}
Daphne Ippolito, Daniel Duckworth, Chris Callison-Burch, and Douglas Eck. 2020.
\newblock \href {https://doi.org/10.18653/v1/2020.acl-main.164} {{Automatic Detection of Generated Text is Easiest when Humans are Fooled}}.
\newblock In \emph{Proceedings of the 58th Annual Meeting of the Association for Computational Linguistics}, pages 1808--1822, Online. Association for Computational Linguistics.

\bibitem[{Jain et~al.(2025)Jain, Synnaeve, and Rozi{\`{e}}re}]{DBLP:conf/iclr/JainSR25}
Kush Jain, Gabriel Synnaeve, and Baptiste Rozi{\`{e}}re. 2025.
\newblock \href {https://openreview.net/forum?id=7o6SG5gVev} {{TestGenEval}: {A} real world unit test generation and test completion benchmark}.
\newblock In \emph{The Thirteenth International Conference on Learning Representations, {ICLR} 2025, Singapore, April 24-28, 2025}. OpenReview.net.

\bibitem[{Jawahar et~al.(2020)Jawahar, Abdul{-}Mageed, and Lakshmanan}]{DBLP:conf/coling/JawaharASL20}
Ganesh Jawahar, Muhammad Abdul{-}Mageed, and Laks V.~S. Lakshmanan. 2020.
\newblock \href {https://doi.org/10.18653/V1/2020.COLING-MAIN.208} {{Automatic Detection of Machine Generated Text: {A} Critical Survey}}.
\newblock In \emph{Proceedings of the 28th International Conference on Computational Linguistics, {COLING} 2020, Barcelona, Spain (Online), December 8-13, 2020}, pages 2296--2309. International Committee on Computational Linguistics.

\bibitem[{Ji et~al.(2024)Ji, Jun, Wu, and Gelles}]{ji2024cybersecurity}
Jessica Ji, Jenny Jun, Maggie Wu, and Rebecca Gelles. 2024.
\newblock \href {https://doi.org/10.51593/2023CA010} {{Cybersecurity Risks of AI-Generated Code}}.
\newblock Technical report, Center for Security and Emerging Technology.
\newblock Center for Security and Emerging Technology.

\bibitem[{JianWang et~al.(2024)JianWang, Liu, Xie, and Li}]{aigc}
JianWang, Shangqing Liu, Xiaofei Xie, and Yi~Li. 2024.
\newblock \href {https://doi.org/10.1145/3691620.3695468} {{An Empirical Study to Evaluate AIGC Detectors on Code Content}}.
\newblock In \emph{Proceedings of the 39th IEEE/ACM International Conference on Automated Software Engineering}, ASE '24, page 844–856, New York, NY, USA. Association for Computing Machinery.

\bibitem[{Katzy et~al.(2025)Katzy, Popescu, van Deursen, and Izadi}]{heap}
Jonathan Katzy, Razvan~Mihai Popescu, Arie van Deursen, and Maliheh Izadi. 2025.
\newblock \href {https://arxiv.org/abs/2501.09653} {{The Heap: A Contamination-Free Multilingual Code Dataset for Evaluating Large Language Models}}.
\newblock \emph{Preprint}, arXiv:2501.09653.

\bibitem[{Khosla et~al.(2020)Khosla, Teterwak, Wang, Sarna, Tian, Isola, Maschinot, Liu, and Krishnan}]{DBLP:conf/nips/KhoslaTWSTIMLK20}
Prannay Khosla, Piotr Teterwak, Chen Wang, Aaron Sarna, Yonglong Tian, Phillip Isola, Aaron Maschinot, Ce~Liu, and Dilip Krishnan. 2020.
\newblock \href {https://proceedings.neurips.cc/paper/2020/hash/d89a66c7c80a29b1bdbab0f2a1a94af8-Abstract.html} {{Supervised Contrastive Learning}}.
\newblock In \emph{Advances in Neural Information Processing Systems 33: Annual Conference on Neural Information Processing Systems 2020, NeurIPS 2020, December 6-12, 2020, virtual}.

\bibitem[{Kipf and Welling(2017)}]{DBLP:conf/iclr/KipfW17}
Thomas~N. Kipf and Max Welling. 2017.
\newblock \href {https://openreview.net/forum?id=SJU4ayYgl} {{Semi-Supervised Classification with Graph Convolutional Networks}}.
\newblock In \emph{5th International Conference on Learning Representations, {ICLR} 2017, Toulon, France, April 24-26, 2017, Conference Track Proceedings}. OpenReview.net.

\bibitem[{Koike et~al.(2024)Koike, Kaneko, and Okazaki}]{DBLP:conf/aaai/KoikeKO24}
Ryuto Koike, Masahiro Kaneko, and Naoaki Okazaki. 2024.
\newblock \href {https://doi.org/10.1609/AAAI.V38I19.30120} {{OUTFOX:} {LLM-Generated Essay Detection Through In-Context Learning with Adversarially Generated Examples}}.
\newblock In \emph{Thirty-Eighth {AAAI} Conference on Artificial Intelligence, {AAAI} 2024, Thirty-Sixth Conference on Innovative Applications of Artificial Intelligence, {IAAI} 2024, Fourteenth Symposium on Educational Advances in Artificial Intelligence, {EAAI} 2014, February 20-27, 2024, Vancouver, Canada}, pages 21258--21266. {AAAI} Press.

\bibitem[{Laugel et~al.(2019)Laugel, Lesot, Marsala, Renard, and Detyniecki}]{posthoc_dangers}
Thibault Laugel, Marie{-}Jeanne Lesot, Christophe Marsala, Xavier Renard, and Marcin Detyniecki. 2019.
\newblock \href {https://doi.org/10.24963/IJCAI.2019/388} {The dangers of post-hoc interpretability: Unjustified counterfactual explanations}.
\newblock In \emph{Proceedings of the Twenty-Eighth International Joint Conference on Artificial Intelligence, {IJCAI} 2019, Macao, China, August 10-16, 2019}, pages 2801--2807. ijcai.org.

\bibitem[{Li et~al.(2023{\natexlab{a}})Li, Su, Chen, Li, and Zhang}]{DBLP:conf/nips/LiSCLZ23}
Hongxin Li, Jingran Su, Yuntao Chen, Qing Li, and Zhaoxiang Zhang. 2023{\natexlab{a}}.
\newblock \href {http://papers.nips.cc/paper\_files/paper/2023/hash/0ff30c4bf31db0119a6219e0d250e037-Abstract-Conference.html} {{{SheetCopilot}: Bringing Software Productivity to the Next Level through Large Language Models}}.
\newblock In \emph{Advances in Neural Information Processing Systems 36: Annual Conference on Neural Information Processing Systems 2023, NeurIPS 2023, New Orleans, LA, USA, December 10 - 16, 2023}.

\bibitem[{Li et~al.(2023{\natexlab{b}})Li, Hong, Fu, Zhang, and Liu}]{DBLP:conf/issre/LiHFZL23}
Ke~Li, Sheng Hong, Cai Fu, Yunhe Zhang, and Ming Liu. 2023{\natexlab{b}}.
\newblock \href {https://doi.org/10.1109/ISSREW60843.2023.00059} {{Discriminating Human-authored from ChatGPT-Generated Code Via Discernable Feature Analysis}}.
\newblock In \emph{34th {IEEE} International Symposium on Software Reliability Engineering, {ISSRE} 2023 - Workshops, Florence, Italy, October 9-12, 2023}, pages 120--127. {IEEE}.

\bibitem[{Li et~al.(2023{\natexlab{c}})Li, Allal, Zi, Muennighoff, Kocetkov, Mou, Marone, Akiki, Li, Chim, Liu, Zheltonozhskii, Zhuo, Wang, Dehaene, Davaadorj, Lamy{-}Poirier, Monteiro, Shliazhko, Gontier, Meade, Zebaze, Yee, Umapathi, Zhu, Lipkin, Oblokulov, Wang, V, Stillerman, Patel, Abulkhanov, Zocca, Dey, Zhang, Fahmy, Bhattacharyya, Yu, Singh, Luccioni, Villegas, Kunakov, Zhdanov, Romero, Lee, Timor, Ding, Schlesinger, Schoelkopf, Ebert, Dao, Mishra, Gu, Robinson, Anderson, Dolan{-}Gavitt, Contractor, Reddy, Fried, Bahdanau, Jernite, Ferrandis, Hughes, Wolf, Guha, von Werra, and de~Vries}]{li2023starcodersourceyou}
Raymond Li, Loubna~Ben Allal, Yangtian Zi, Niklas Muennighoff, Denis Kocetkov, Chenghao Mou, Marc Marone, Christopher Akiki, Jia Li, Jenny Chim, Qian Liu, Evgenii Zheltonozhskii, Terry~Yue Zhuo, Thomas Wang, Olivier Dehaene, Mishig Davaadorj, Joel Lamy{-}Poirier, Jo{\~{a}}o Monteiro, Oleh Shliazhko, and 48 others. 2023{\natexlab{c}}.
\newblock \href {https://openreview.net/forum?id=KoFOg41haE} {Starcoder: may the source be with you!}
\newblock \emph{Trans. Mach. Learn. Res.}, 2023.

\bibitem[{Li et~al.(2023{\natexlab{d}})Li, Fu, Zhang, Huang, Sun, Lyu, Liu, Jin, and Li}]{taco}
Rongao Li, Jie Fu, Bo-Wen Zhang, Tao Huang, Zhihong Sun, Chen Lyu, Guang Liu, Zhi Jin, and Ge~Li. 2023{\natexlab{d}}.
\newblock {{TACO}: Topics in Algorithmic COde generation dataset}.
\newblock \emph{arXiv preprint arXiv:2312.14852}.

\bibitem[{Li and Li(2024)}]{DBLP:conf/acl/LiL24}
Xianming Li and Jing Li. 2024.
\newblock \href {https://doi.org/10.18653/V1/2024.ACL-LONG.101} {{{AoE}: Angle-optimized Embeddings for Semantic Textual Similarity}}.
\newblock In \emph{Proceedings of the 62nd Annual Meeting of the Association for Computational Linguistics (Volume 1: Long Papers), {ACL} 2024, Bangkok, Thailand, August 11-16, 2024}, pages 1825--1839. Association for Computational Linguistics.

\bibitem[{Li et~al.(2024)Li, Li, Cui, Bi, Wang, Wang, Yang, Shi, and Zhang}]{mage}
Yafu Li, Qintong Li, Leyang Cui, Wei Bi, Zhilin Wang, Longyue Wang, Linyi Yang, Shuming Shi, and Yue Zhang. 2024.
\newblock \href {https://doi.org/10.18653/v1/2024.acl-long.3} {{MAGE}: {Machine-generated Text Detection in the Wild}}.
\newblock In \emph{Proceedings of the 62nd Annual Meeting of the Association for Computational Linguistics (Volume 1: Long Papers)}, pages 36--53, Bangkok, Thailand. Association for Computational Linguistics.

\bibitem[{Loshchilov and Hutter(2019)}]{adamw}
Ilya Loshchilov and Frank Hutter. 2019.
\newblock \href {https://openreview.net/forum?id=Bkg6RiCqY7} {{Decoupled Weight Decay Regularization}}.
\newblock In \emph{7th International Conference on Learning Representations, {ICLR} 2019, New Orleans, LA, USA, May 6-9, 2019}. OpenReview.net.

\bibitem[{Lozhkov et~al.()Lozhkov, Li, Allal, Cassano, Lamy-Poirier, Tazi, Tang, Pykhtar, Liu, Wei, Liu, Tian, Kocetkov, Zucker, Belkada, Wang, Liu, Abulkhanov, Paul, Li, Li, Risdal, Li, Zhu, Zhuo, Zheltonozhskii, Dade, Yu, Krauß, Jain, Su, He, Dey, Abati, Chai, Muennighoff, Tang, Oblokulov, Akiki, Marone, Mou, Mishra, Gu, Hui, Dao, Zebaze, Dehaene, Patry, Xu, McAuley, Hu, Scholak, Paquet, Robinson, Anderson, Chapados, Patwary, Tajbakhsh, Jernite, Ferrandis, Zhang, Hughes, Wolf, Guha, von Werra, and de~Vries}]{lozhkov2024starcoder}
Anton Lozhkov, Raymond Li, Loubna~Ben Allal, Federico Cassano, Joel Lamy-Poirier, Nouamane Tazi, Ao~Tang, Dmytro Pykhtar, Jiawei Liu, Yuxiang Wei, Tianyang Liu, Max Tian, Denis Kocetkov, Arthur Zucker, Younes Belkada, Zijian Wang, Qian Liu, Dmitry Abulkhanov, Indraneil Paul, and 47 others.
\newblock \href {https://arxiv.org/abs/2402.19173} {{StarCoder 2 and The Stack v2: The Next Generation}}.
\newblock \emph{Preprint}, arXiv:2402.19173.

\bibitem[{Lu et~al.(2025)Lu, Zhou, Wang, Ren, Shi, Pan, Zhan, and Li}]{DBLP:conf/iclr/LuZ0RSPZL25}
Zimu Lu, Aojun Zhou, Ke~Wang, Houxing Ren, Weikang Shi, Junting Pan, Mingjie Zhan, and Hongsheng Li. 2025.
\newblock \href {https://openreview.net/forum?id=1Iuw1jcIrf} {{MathCoder2}: Better math reasoning from continued pretraining on model-translated mathematical code}.
\newblock In \emph{The Thirteenth International Conference on Learning Representations, {ICLR} 2025, Singapore, April 24-28, 2025}. OpenReview.net.

\bibitem[{Lundberg and Lee(2017)}]{shap}
Scott~M. Lundberg and Su-In Lee. 2017.
\newblock A unified approach to interpreting model predictions.
\newblock In \emph{Proceedings of the 31st International Conference on Neural Information Processing Systems}, NIPS'17, page 4768–4777, Red Hook, NY, USA. Curran Associates Inc.

\bibitem[{Lyu et~al.(2024)Lyu, Apidianaki, and Callison{-}Burch}]{DBLP:journals/coling/LyuAC24}
Qing Lyu, Marianna Apidianaki, and Chris Callison{-}Burch. 2024.
\newblock \href {https://doi.org/10.1162/COLI\_A\_00511} {Towards faithful model explanation in {NLP:} {A} survey}.
\newblock \emph{Comput. Linguistics}, 50(2):657--723.

\bibitem[{Macko et~al.(2023)Macko, M{\'{o}}ro, Uchendu, Lucas, Yamashita, Pikuliak, Srba, Le, Lee, Simko, and Bielikov{\'{a}}}]{multitude}
Dominik Macko, R{\'{o}}bert M{\'{o}}ro, Adaku Uchendu, Jason~Samuel Lucas, Michiharu Yamashita, Mat{\'{u}}s Pikuliak, Ivan Srba, Thai Le, Dongwon Lee, Jakub Simko, and M{\'{a}}ria Bielikov{\'{a}}. 2023.
\newblock \href {https://doi.org/10.48550/ARXIV.2310.13606} {{{MULTITuDE}: Large-Scale Multilingual Machine-Generated Text Detection Benchmark}}.
\newblock \emph{CoRR}, abs/2310.13606.

\bibitem[{Manh et~al.(2023)Manh, Hai, Dau, Nguyen, Nghiem, Guo, and Bui}]{thevault}
Dung~Nguyen Manh, Le~Nam Hai, Anh T.~V. Dau, Anh~Minh Nguyen, Khanh Nghiem, Jin Guo, and Nghi D.~Q. Bui. 2023.
\newblock \href {https://doi.org/10.18653/V1/2023.FINDINGS-EMNLP.316} {{The Vault: {A} Comprehensive Multilingual Dataset for Advancing Code Understanding and Generation}}.
\newblock In \emph{Findings of the Association for Computational Linguistics: {EMNLP} 2023, Singapore, December 6-10, 2023}, pages 4763--4788. Association for Computational Linguistics.

\bibitem[{Martini(2015)}]{DBLP:conf/hapoc/Martini15}
Simone Martini. 2015.
\newblock \href {https://doi.org/10.1007/978-3-319-47286-7\_15} {{Several Types of Types in Programming Languages}}.
\newblock In \emph{History and Philosophy of Computing - Third International Conference, HaPoC 2015, Pisa, Italy, October 8-11, 2015, Revised Selected Papers}, volume 487 of \emph{{IFIP} Advances in Information and Communication Technology}, pages 216--227.

\bibitem[{Masrour et~al.(2025)Masrour, Emi, and Spero}]{DBLP:journals/corr/abs-2501-03437}
Elyas Masrour, Bradley Emi, and Max Spero. 2025.
\newblock \href {https://doi.org/10.48550/ARXIV.2501.03437} {{{DAMAGE:} Detecting Adversarially Modified {AI} Generated Text}}.
\newblock \emph{CoRR}, abs/2501.03437.

\bibitem[{Mishra et~al.(2024)Mishra, Stallone, Zhang, Shen, Prasad, Soria, Merler, Selvam, Surendran, Singh, Sethi, Dang, Li, Wu, Zawad, Coleman, White, Lewis, Pavuluri, Koyfman, Lublinsky, de~Bayser, Abdelaziz, Basu, Agarwal, Zhou, Johnson, Goyal, Patel, Shah, Zerfos, Ludwig, Munawar, Crouse, Kapanipathi, Salaria, Calio, Wen, Seelam, Belgodere, Fonseca, Singhee, Desai, Cox, Puri, and Panda}]{mishra2024granitecodemodelsfamily}
Mayank Mishra, Matt Stallone, Gaoyuan Zhang, Yikang Shen, Aditya Prasad, Adriana~Meza Soria, Michele Merler, Parameswaran Selvam, Saptha Surendran, Shivdeep Singh, Manish Sethi, Xuan-Hong Dang, Pengyuan Li, Kun-Lung Wu, Syed Zawad, Andrew Coleman, Matthew White, Mark Lewis, Raju Pavuluri, and 27 others. 2024.
\newblock \href {https://arxiv.org/abs/2405.04324} {{Granite Code Models: A Family of Open Foundation Models for Code Intelligence}}.
\newblock \emph{Preprint}, arXiv:2405.04324.

\bibitem[{Moll{\'a} et~al.(2024)Moll{\'a}, Xu, Zeng, and Li}]{alta}
Diego Moll{\'a}, Qiongkai Xu, Zijie Zeng, and Zhuang Li. 2024.
\newblock \href {https://aclanthology.org/2024.alta-1.17/} {{Overview of the 2024 {ALTA} Shared Task: Detect Automatic {AI}-Generated Sentences for Human-{AI} Hybrid Articles}}.
\newblock In \emph{Proceedings of the 22nd Annual Workshop of the Australasian Language Technology Association}, pages 197--202, Canberra, Australia. Association for Computational Linguistics.

\bibitem[{Murali et~al.(2024)Murali, Maddila, Ahmad, Bolin, Cheng, Ghorbani, Fernandez, Nagappan, and Rigby}]{DBLP:journals/pacmse/MuraliMABCGFNR24}
Vijayaraghavan Murali, Chandra~Shekhar Maddila, Imad Ahmad, Michael Bolin, Daniel Cheng, Negar Ghorbani, Renuka Fernandez, Nachiappan Nagappan, and Peter~C. Rigby. 2024.
\newblock \href {https://doi.org/10.1145/3643774} {{AI-Assisted Code Authoring at Scale: Fine-Tuning, Deploying, and Mixed Methods Evaluation}}.
\newblock \emph{Proc. {ACM} Softw. Eng.}, 1({FSE}):1066--1085.

\bibitem[{Nguyen et~al.(2024)Nguyen, {Di Rocco}, {Di Sipio}, Rubei, {Di Ruscio}, and {Di Penta}}]{NGUYEN2024112059}
Phuong~T. Nguyen, Juri {Di Rocco}, Claudio {Di Sipio}, Riccardo Rubei, Davide {Di Ruscio}, and Massimiliano {Di Penta}. 2024.
\newblock \href {https://doi.org/10.1016/j.jss.2024.112059} {{GPTSniffer: A CodeBERT-based classifier to detect source code written by ChatGPT}}.
\newblock \emph{Journal of Systems and Software}, page 112059.

\bibitem[{Nunes et~al.(2025)Nunes, Figueiredo, Soares, Nadi, Ferreira, and dos Santos}]{DBLP:journals/corr/abs-2502-02368}
Henrique~Gomes Nunes, Eduardo Figueiredo, Larissa~Rocha Soares, Sarah Nadi, Fischer Ferreira, and Geanderson~E. dos Santos. 2025.
\newblock \href {https://doi.org/10.1109/SANER64311.2025.00069} {{Evaluating the Effectiveness of LLMs in Fixing Maintainability Issues in Real-World Projects}}.
\newblock In \emph{{IEEE} International Conference on Software Analysis, Evolution and Reengineering, {SANER} 2025, Montreal, QC, Canada, March 4-7, 2025}, pages 669--680. {IEEE}.

\bibitem[{Orel et~al.(2025)Orel, Azizov, and Nakov}]{orel2025codetm4detectingmachinegeneratedcode}
Daniil Orel, Dilshod Azizov, and Preslav Nakov. 2025.
\newblock \href {https://arxiv.org/abs/2503.13733} {{CoDet-M4: Detecting Machine-Generated Code in Multi-Lingual, Multi-Generator and Multi-Domain Settings}}.
\newblock \emph{Preprint}, arXiv:2503.13733.

\bibitem[{Pan et~al.(2024)Pan, Chok, Wong, Shin, Poon, Yang, Chong, Lo, and Lim}]{detectors}
Wei~Hung Pan, Ming~Jie Chok, Jonathan Leong~Shan Wong, Yung~Xin Shin, Yeong~Shian Poon, Zhou Yang, Chun~Yong Chong, David Lo, and Mei~Kuan Lim. 2024.
\newblock \href {https://doi.org/10.1145/3639474.3640068} {{Assessing AI Detectors in Identifying AI-Generated Code: Implications for Education}}.
\newblock In \emph{Proceedings of the 46th International Conference on Software Engineering: Software Engineering Education and Training}, ICSE-SEET '24, page 1–11, New York, NY, USA. Association for Computing Machinery.

\bibitem[{Paul et~al.(2025)Paul, Yang, Glavas, Kersting, and Gurevych}]{obscura}
Indraneil Paul, Haoyi Yang, Goran Glavas, Kristian Kersting, and Iryna Gurevych. 2025.
\newblock \href {https://openreview.net/forum?id=VYvxrD7aS0} {{ObscuraCoder: Powering Efficient Code {LM} Pre-Training Via Obfuscation Grounding}}.
\newblock In \emph{The Thirteenth International Conference on Learning Representations, {ICLR} 2025, Singapore, April 24-28, 2025}. OpenReview.net.

\bibitem[{Pearce et~al.(2025)Pearce, Ahmad, Tan, Dolan{-}Gavitt, and Karri}]{DBLP:journals/cacm/PearceATDK25}
Hammond Pearce, Baleegh Ahmad, Benjamin Tan, Brendan Dolan{-}Gavitt, and Ramesh Karri. 2025.
\newblock \href {https://doi.org/10.1145/3610721} {{Asleep at the Keyboard? Assessing the Security of GitHub Copilot's Code Contributions}}.
\newblock \emph{Commun. {ACM}}, 68(2):96--105.

\bibitem[{Prokhorenkova et~al.(2018)Prokhorenkova, Gusev, Vorobev, Dorogush, and Gulin}]{catboost}
Liudmila Prokhorenkova, Gleb Gusev, Aleksandr Vorobev, Anna~Veronika Dorogush, and Andrey Gulin. 2018.
\newblock \href {https://proceedings.neurips.cc/paper_files/paper/2018/file/14491b756b3a51daac41c24863285549-Paper.pdf} {{CatBoost}: unbiased boosting with categorical features}.
\newblock In \emph{Advances in Neural Information Processing Systems}, volume~31. Curran Associates, Inc.

\bibitem[{Puri et~al.(2021)Puri, Kung, Janssen, Zhang, Domeniconi, Zolotov, Dolby, Chen, Choudhury, Decker, Thost, Buratti, Pujar, Ramji, Finkler, Malaika, and Reiss}]{codenet}
Ruchir Puri, David~S. Kung, Geert Janssen, Wei Zhang, Giacomo Domeniconi, Vladimir Zolotov, Julian Dolby, Jie Chen, Mihir Choudhury, Lindsey Decker, Veronika Thost, Luca Buratti, Saurabh Pujar, Shyam Ramji, Ulrich Finkler, Susan Malaika, and Frederick Reiss. 2021.
\newblock \href {https://arxiv.org/abs/2105.12655} {{CodeNet}: A large-scale ai for code dataset for learning a diversity of coding tasks}.
\newblock \emph{Preprint}, arXiv:2105.12655.

\bibitem[{Puryear and Sprint(2022)}]{moss_copilot}
Ben Puryear and Gina Sprint. 2022.
\newblock {Github copilot in the classroom: learning to code with AI assistance}.
\newblock \emph{J. Comput. Sci. Coll.}, 38(1):37–47.

\bibitem[{Qwen et~al.(2025)Qwen, :, Yang, Yang, Zhang, Hui, Zheng, Yu, Li, Liu, Huang, Wei, Lin, Yang, Tu, Zhang, Yang, Yang, Zhou, Lin, Dang, Lu, Bao, Yang, Yu, Li, Xue, Zhang, Zhu, Men, Lin, Li, Tang, Xia, Ren, Ren, Fan, Su, Zhang, Wan, Liu, Cui, Zhang, and Qiu}]{qwen}
Qwen, :, An~Yang, Baosong Yang, Beichen Zhang, Binyuan Hui, Bo~Zheng, Bowen Yu, Chengyuan Li, Dayiheng Liu, Fei Huang, Haoran Wei, Huan Lin, Jian Yang, Jianhong Tu, Jianwei Zhang, Jianxin Yang, Jiaxi Yang, Jingren Zhou, and 25 others. 2025.
\newblock \href {https://arxiv.org/abs/2412.15115} {Qwen2.5 technical report}.
\newblock \emph{Preprint}, arXiv:2412.15115.

\bibitem[{Rafailov et~al.(2023)Rafailov, Sharma, Mitchell, Manning, Ermon, and Finn}]{dpo}
Rafael Rafailov, Archit Sharma, Eric Mitchell, Christopher~D Manning, Stefano Ermon, and Chelsea Finn. 2023.
\newblock \href {https://proceedings.neurips.cc/paper_files/paper/2023/file/a85b405ed65c6477a4fe8302b5e06ce7-Paper-Conference.pdf} {{Direct Preference Optimization: Your Language Model is Secretly a Reward Model}}.
\newblock In \emph{Advances in Neural Information Processing Systems}, volume~36, pages 53728--53741. Curran Associates, Inc.

\bibitem[{Sadasivan et~al.(2023)Sadasivan, Kumar, Balasubramanian, Wang, and Feizi}]{can_ai_generated_text_be_detected}
Vinu~Sankar Sadasivan, Aounon Kumar, Sriram Balasubramanian, Wenxiao Wang, and Soheil Feizi. 2023.
\newblock \href {https://doi.org/10.48550/ARXIV.2303.11156} {{Can AI-Generated Text be Reliably Detected?}}
\newblock \emph{CoRR}, abs/2303.11156.

\bibitem[{Schulman et~al.(2017)Schulman, Wolski, Dhariwal, Radford, and Klimov}]{ppo}
John Schulman, Filip Wolski, Prafulla Dhariwal, Alec Radford, and Oleg Klimov. 2017.
\newblock \href {https://arxiv.org/abs/1707.06347} {{Proximal Policy Optimization Algorithms}}.
\newblock \emph{CoRR}, abs/1707.06347.

\bibitem[{Shamardina et~al.(2022)Shamardina, Mikhailov, Chernianskii, Fenogenova, Saidov, Valeeva, Shavrina, Smurov, Tutubalina, and Artemova}]{ruatd}
Tatiana Shamardina, Vladislav Mikhailov, Daniil Chernianskii, Alena Fenogenova, Marat Saidov, Anastasiya Valeeva, Tatiana Shavrina, Ivan Smurov, Elena Tutubalina, and Ekaterina Artemova. 2022.
\newblock \href {https://doi.org/10.28995/2075-7182-2022-21-497-511} {{Findings of the The RuATD Shared Task 2022 on Artificial Text Detection in Russian}}.
\newblock In \emph{Computational Linguistics and Intellectual Technologies}, page 497–511. RSUH.

\bibitem[{Shao et~al.(2024)Shao, Wang, Zhu, Xu, Song, Zhang, Li, Wu, and Guo}]{grpo}
Zhihong Shao, Peiyi Wang, Qihao Zhu, Runxin Xu, Junxiao Song, Mingchuan Zhang, Y.~K. Li, Y.~Wu, and Daya Guo. 2024.
\newblock \href {https://doi.org/10.48550/ARXIV.2402.03300} {{DeepSeekMath: Pushing the Limits of Mathematical Reasoning in Open Language Models}}.
\newblock \emph{CoRR}, abs/2402.03300.

\bibitem[{Shi et~al.(2024)Shi, Wang, Yin, Chen, Chang, and Hsieh}]{red_teaming}
Zhouxing Shi, Yihan Wang, Fan Yin, Xiangning Chen, Kai-Wei Chang, and Cho-Jui Hsieh. 2024.
\newblock \href {https://doi.org/10.1162/tacl_a_00639} {{Red Teaming Language Model Detectors with Language Models}}.
\newblock \emph{Transactions of the Association for Computational Linguistics}, 12:174--189.

\bibitem[{Shumailov et~al.(2024)Shumailov, Shumaylov, Zhao, Papernot, Anderson, and Gal}]{DBLP:journals/nature/ShumailovSZPAG24}
Ilia Shumailov, Zakhar Shumaylov, Yiren Zhao, Nicolas Papernot, Ross~J. Anderson, and Yarin Gal. 2024.
\newblock \href {https://doi.org/10.1038/S41586-024-07566-Y} {{AI} models collapse when trained on recursively generated data}.
\newblock \emph{Nat.}, 631(8022):755--759.

\bibitem[{Su et~al.(2024)Su, Kong, Lin, Jennings, Norick, Kliegl, Patwary, Shoeybi, and Catanzaro}]{DBLP:journals/corr/abs-2412-02595}
Dan Su, Kezhi Kong, Ying Lin, Joseph Jennings, Brandon Norick, Markus Kliegl, Mostofa Patwary, Mohammad Shoeybi, and Bryan Catanzaro. 2024.
\newblock \href {https://doi.org/10.48550/ARXIV.2412.02595} {{{Nemotron-CC}: Transforming Common Crawl into a Refined Long-Horizon Pretraining Dataset}}.
\newblock \emph{CoRR}, abs/2412.02595.

\bibitem[{Su et~al.(2023)Su, Wu, Zhou, Ma, and Hu}]{hc3_plus}
Zhenpeng Su, Xing Wu, Wei Zhou, Guangyuan Ma, and Songlin Hu. 2023.
\newblock \href {https://doi.org/10.48550/arXiv.2309.02731} {{HC3 Plus: A Semantic-Invariant Human ChatGPT Comparison Corpus}}.
\newblock \emph{CoRR}, abs/2309.02731.

\bibitem[{Tulchinskii et~al.(2023)Tulchinskii, Kuznetsov, Kushnareva, Cherniavskii, Nikolenko, Burnaev, Barannikov, and Piontkovskaya}]{DBLP:conf/nips/TulchinskiiKKCN23}
Eduard Tulchinskii, Kristian Kuznetsov, Laida Kushnareva, Daniil Cherniavskii, Sergey~I. Nikolenko, Evgeny Burnaev, Serguei Barannikov, and Irina Piontkovskaya. 2023.
\newblock \href {http://papers.nips.cc/paper\_files/paper/2023/hash/7baa48bc166aa2013d78cbdc15010530-Abstract-Conference.html} {{Intrinsic Dimension Estimation for Robust Detection of AI-Generated Texts}}.
\newblock In \emph{Advances in Neural Information Processing Systems 36: Annual Conference on Neural Information Processing Systems 2023, NeurIPS 2023, New Orleans, LA, USA, December 10 - 16, 2023}.

\bibitem[{Vaithilingam et~al.(2022)Vaithilingam, Zhang, and Glassman}]{copilot_obstacles}
Priyan Vaithilingam, Tianyi Zhang, and Elena~L. Glassman. 2022.
\newblock \href {https://doi.org/10.1145/3491101.3519665} {{Expectation vs. Experience: Evaluating the Usability of Code Generation Tools Powered by Large Language Models}}.
\newblock In \emph{Extended Abstracts of the 2022 CHI Conference on Human Factors in Computing Systems}, CHI EA '22, New York, NY, USA. Association for Computing Machinery.

\bibitem[{Veselovsky et~al.(2023)Veselovsky, Ribeiro, and West}]{DBLP:journals/corr/abs-2306-07899}
Veniamin Veselovsky, Manoel~Horta Ribeiro, and Robert West. 2023.
\newblock \href {https://doi.org/10.48550/ARXIV.2306.07899} {{Artificial Artificial Artificial Intelligence: Crowd Workers Widely Use Large Language Models for Text Production Tasks}}.
\newblock \emph{CoRR}, abs/2306.07899.

\bibitem[{Wang et~al.(2022)Wang, Yang, Huang, Jiao, Yang, Jiang, Majumder, and Wei}]{DBLP:journals/corr/abs-2212-03533}
Liang Wang, Nan Yang, Xiaolong Huang, Binxing Jiao, Linjun Yang, Daxin Jiang, Rangan Majumder, and Furu Wei. 2022.
\newblock \href {https://doi.org/10.48550/ARXIV.2212.03533} {{Text Embeddings by Weakly-Supervised Contrastive Pre-training}}.
\newblock \emph{CoRR}, abs/2212.03533.

\bibitem[{Wang et~al.(2024)Wang, Mansurov, Ivanov, Su, Shelmanov, Tsvigun, Whitehouse, Mohammed~Afzal, Mahmoud, Sasaki, Arnold, Aji, Habash, Gurevych, and Nakov}]{m4}
Yuxia Wang, Jonibek Mansurov, Petar Ivanov, Jinyan Su, Artem Shelmanov, Akim Tsvigun, Chenxi Whitehouse, Osama Mohammed~Afzal, Tarek Mahmoud, Toru Sasaki, Thomas Arnold, Alham~Fikri Aji, Nizar Habash, Iryna Gurevych, and Preslav Nakov. 2024.
\newblock \href {https://aclanthology.org/2024.eacl-long.83/} {{M4: Multi-generator, Multi-domain, and Multi-lingual Black-Box Machine-Generated Text Detection}}.
\newblock In \emph{Proceedings of the 18th Conference of the European Chapter of the Association for Computational Linguistics (Volume 1: Long Papers)}, pages 1369--1407, St. Julian{'}s, Malta. Association for Computational Linguistics.

\bibitem[{Wang et~al.(2025)Wang, Shelmanov, Mansurov, Tsvigun, Mikhailov, Xing, Xie, Geng, Puccetti, Artemova, Su, Ta, Abassy, Elozeiri, El~Etter, Goloburda, Mahmoud, Tomar, Laiyk, Mohammed~Afzal, Koike, Kaneko, Aji, Habash, Gurevych, and Nakov}]{wang-etal-2025-genai}
Yuxia Wang, Artem Shelmanov, Jonibek Mansurov, Akim Tsvigun, Vladislav Mikhailov, Rui Xing, Zhuohan Xie, Jiahui Geng, Giovanni Puccetti, Ekaterina Artemova, Jinyan Su, Minh~Ngoc Ta, Mervat Abassy, Kareem~Ashraf Elozeiri, Saad El Dine~Ahmed El~Etter, Maiya Goloburda, Tarek Mahmoud, Raj~Vardhan Tomar, Nurkhan Laiyk, and 7 others. 2025.
\newblock \href {https://aclanthology.org/2025.genaidetect-1.27/} {{{G}en{AI} Content Detection Task 1: {E}nglish and Multilingual Machine-Generated Text Detection: {AI} vs. Human}}.
\newblock In \emph{Proceedings of the 1stWorkshop on GenAI Content Detection (GenAIDetect)}, pages 244--261, Abu Dhabi, UAE. International Conference on Computational Linguistics.

\bibitem[{Warner et~al.(2024)Warner, Chaffin, Clavi{\'{e}}, Weller, Hallstr{\"{o}}m, Taghadouini, Gallagher, Biswas, Ladhak, Aarsen, Cooper, Adams, Howard, and Poli}]{modernbert}
Benjamin Warner, Antoine Chaffin, Benjamin Clavi{\'{e}}, Orion Weller, Oskar Hallstr{\"{o}}m, Said Taghadouini, Alexis Gallagher, Raja Biswas, Faisal Ladhak, Tom Aarsen, Nathan Cooper, Griffin Adams, Jeremy Howard, and Iacopo Poli. 2024.
\newblock \href {https://doi.org/10.48550/ARXIV.2412.13663} {{Smarter, Better, Faster, Longer: {A} Modern Bidirectional Encoder for Fast, Memory Efficient, and Long Context Finetuning and Inference}}.
\newblock \emph{CoRR}, abs/2412.13663.

\bibitem[{Weber et~al.(2024{\natexlab{a}})Weber, Fu, Anthony, Oren, Adams, Alexandrov, Lyu, Nguyen, Yao, Adams, Athiwaratkun, Chalamala, Chen, Ryabinin, Dao, Liang, R{\'{e}}, Rish, and Zhang}]{DBLP:conf/nips/WeberFAOAALNYAA24}
Maurice Weber, Daniel~Y. Fu, Quentin Anthony, Yonatan Oren, Shane Adams, Anton Alexandrov, Xiaozhong Lyu, Huu Nguyen, Xiaozhe Yao, Virginia Adams, Ben Athiwaratkun, Rahul Chalamala, Kezhen Chen, Max Ryabinin, Tri Dao, Percy Liang, Christopher R{\'{e}}, Irina Rish, and Ce~Zhang. 2024{\natexlab{a}}.
\newblock \href {http://papers.nips.cc/paper\_files/paper/2024/hash/d34497330b1fd6530f7afd86d0df9f76-Abstract-Datasets\_and\_Benchmarks\_Track.html} {{RedPajama: an Open Dataset for Training Large Language Models}}.
\newblock In \emph{Advances in Neural Information Processing Systems 38: Annual Conference on Neural Information Processing Systems 2024, NeurIPS 2024, Vancouver, BC, Canada, December 10 - 15, 2024}.

\bibitem[{Weber et~al.(2024{\natexlab{b}})Weber, Brandmaier, Schmidt, and Mayer}]{DBLP:journals/pacmhci/WeberBSM24}
Thomas Weber, Maximilian Brandmaier, Albrecht Schmidt, and Sven Mayer. 2024{\natexlab{b}}.
\newblock \href {https://doi.org/10.1145/3661145} {{Significant Productivity Gains through Programming with Large Language Models}}.
\newblock \emph{Proc. {ACM} Hum. Comput. Interact.}, 8({EICS}):1--29.

\bibitem[{Wu et~al.(2025)Wu, Huang, Shi, Wang, Pu, Gao, Liu, Nan, Yuan, Zhang, Zhang, Du, Guo, Yin, Hu, and Chen}]{inversecoder}
Yutong Wu, Di~Huang, Wenxuan Shi, Wei Wang, Yewen Pu, Lingzhe Gao, Shihao Liu, Ziyuan Nan, Kaizhao Yuan, Rui Zhang, Xishan Zhang, Zidong Du, Qi~Guo, Dawei Yin, Xing Hu, and Yunji Chen. 2025.
\newblock \href {https://doi.org/10.1609/AAAI.V39I24.34742} {Inversecoder: Self-improving instruction-tuned code llms with inverse-instruct}.
\newblock In \emph{AAAI-25, Sponsored by the Association for the Advancement of Artificial Intelligence, February 25 - March 4, 2025, Philadelphia, PA, {USA}}, pages 25525--25533. {AAAI} Press.

\bibitem[{Xie et~al.(2020)Xie, Luong, Hovy, and Le}]{DBLP:conf/cvpr/XieLHL20}
Qizhe Xie, Minh{-}Thang Luong, Eduard~H. Hovy, and Quoc~V. Le. 2020.
\newblock \href {https://doi.org/10.1109/CVPR42600.2020.01070} {{Self-Training With Noisy Student Improves ImageNet Classification}}.
\newblock In \emph{2020 {IEEE/CVF} Conference on Computer Vision and Pattern Recognition, {CVPR} 2020, Seattle, WA, USA, June 13-19, 2020}, pages 10684--10695. Computer Vision Foundation / {IEEE}.

\bibitem[{Xu et~al.(2025)Xu, Ni, Guo, Liu, Wang, Liu, and Yang}]{gptsensor}
Xiaodan Xu, Chao Ni, Xinrong Guo, Shaoxuan Liu, Xiaoya Wang, Kui Liu, and Xiaohu Yang. 2025.
\newblock \href {https://doi.org/10.1145/3705300} {{Distinguishing LLM-Generated from Human-Written Code by Contrastive Learning}}.
\newblock \emph{ACM Trans. Softw. Eng. Methodol.}, 34(4).

\bibitem[{Yalniz et~al.(2019)Yalniz, J{\'{e}}gou, Chen, Paluri, and Mahajan}]{DBLP:journals/corr/abs-1905-00546}
I.~Zeki Yalniz, Herv{\'{e}} J{\'{e}}gou, Kan Chen, Manohar Paluri, and Dhruv Mahajan. 2019.
\newblock \href {https://arxiv.org/abs/1905.00546} {{Billion-scale semi-supervised learning for image classification}}.
\newblock \emph{CoRR}, abs/1905.00546.

\bibitem[{Yang et~al.(2023)Yang, Zhang, Chen, Petzold, Wang, and Cheng}]{DBLP:journals/corr/abs-2310-05103}
Xianjun Yang, Kexun Zhang, Haifeng Chen, Linda~R. Petzold, William~Yang Wang, and Wei Cheng. 2023.
\newblock \href {https://doi.org/10.48550/ARXIV.2310.05103} {{Zero-Shot Detection of Machine-Generated Codes}}.
\newblock \emph{CoRR}, abs/2310.05103.

\end{thebibliography}

\clearpage
\tableofcontents
\appendix

\section{List Of Models Used}
\label{appx:model_names}
\Cref{tab:models} illustrates that we are using a diverse set of models from 11 model families, combining both instruct and base versions of models. We are also covering a diverse set of sizes: from 2B up to 72B, and use both open-weights and API-based models.
\begin{table}[!tbp]
\centering
\scalebox{0.78}{
\begin{tabular}{ll}
\toprule
\textbf{\texttt{Model Family}} & \textbf{\texttt{Model}} \\
\midrule

\multirow{4}{*}{\textbf{\texttt{Yi}}} & \texttt{Yi-Coder-9B} \\
    
& \texttt{Yi-Coder-9B-Chat} \\
    
& \texttt{Yi-Coder-1.5B-Chat} \\
    
& \texttt{Yi-Coder-1.5B}   \\
\midrule
\multirow{2}{*}{\textbf{\texttt{GPT}}} & \texttt{GPT-4o-mini} \\
& \texttt{GPT-4o}  \\
\midrule
\multirow{7}{*}{\textbf{\texttt{Qwen}}} & \texttt{Qwen2.5-Coder-7B} \\
& \texttt{Qwen2.5-Coder-7B-Instruct} \\
& \texttt{Qwen2.5-Coder-1.5B-Instruct} \\
& \texttt{Qwen2.5-Coder-32B-Instruct} \\
& \texttt{Qwen2.5-72B-Instruct} \\
& \texttt{Qwen2.5-Coder-1.5B} \\
& \texttt{Qwen2.5-Coder-14B-Instruct}   \\
\midrule
\multirow{3}{*}{\textbf{\texttt{Gemma}}} & \texttt{codegemma-7b-it} \\
& \texttt{codegemma-7b} \\
& \texttt{codegemma-2b}    \\
\midrule
\multirow{3}{*}{\textbf{\texttt{CodeLlama}}} & CodeLlama-70b-Instruct-hf\\
    
& \texttt{CodeLlama-34b-Instruct-hf} \\
    &\texttt{CodeLlama-7b-hf}
   \\
\midrule
\multirow{4}{*}{\textbf{\texttt{Deepseek}}} & \texttt{deepseek-coder-6.7b-instruct} \\
& \texttt{deepseek-coder-6.7b-base} \\
& \texttt{deepseek-coder-1.3b-instruct} \\
& \texttt{deepseek-coder-1.3b-base} \\
\midrule
\multirow{2}{*}{\textbf{\texttt{Granite}}} & \texttt{granite-8b-code-instruct-4k} \\
& \texttt{granite-8b-code-base-4k}
   \\
\midrule
\multirow{7}{*}{\textbf{\texttt{Llama}}} & \texttt{Llama-3.1-8B-Instruct} \\
    
& \texttt{Llama-3.2-3B} \\
& \texttt{Llama-3.1-70B-Instruct} \\
& \texttt{Llama-3.3-70B-Instruct} \\
& \texttt{Llama-3.3-70B-Instruct-Turbo} \\
& \texttt{Llama-3.2-1B} \\
& \texttt{Llama-3.1-8B} 
   \\
\midrule
\multirow{6}{*}{\textbf{\texttt{Phi}}} & \texttt{Phi-3-small-8k-instruct} \\
& \texttt{Phi-3-mini-4k-instruct} \\
& \texttt{phi-4} \\
& \texttt{Phi-3-medium-4k-instruct} \\
& \texttt{phi-2} \\
& \texttt{Phi-3.5-mini-instruct} \\
\midrule
\textbf{\texttt{Mistral}} & \texttt{Mistral-Small-24B-Instruct-2501}
   \\
\midrule
\multirow{4}{*}{\textbf{\texttt{StarCoder}}} & \texttt{starcoder2-15B} \\
& \texttt{starcoder} \\
& \texttt{starcoder2-7b} \\
& \texttt{starcoder2-3b} \\

\bottomrule
\end{tabular}
}
\caption{Model families and their selected models used in \texttt{DroidCollection}.}
\label{tab:models}
\end{table}

\section{Dataset Creation and Statistics}
\label{appx:data}

\subsection{Inverse Instructions Setup}
\label{appx:inverse}
For inverse instruction creation, we applied 4 LLMs: GPT-4o-mini, Llama3.1 8B, Qwen2.5 7B, and Phi-3 small (7B). These models were given the code, and they were asked to generate their summary and a prompt which could result in an LLM generating it. The prompt is given on \Cref{lst:code-analysis}.
{
\lstset{
  basicstyle=\ttfamily\small,
  frame=single,
  backgroundcolor=\color{gray!5},
  breaklines=true,
}
\begin{lstlisting}[language=Python, caption={Prompt for code analysis and LLM prompt generation}, label={lst:code-analysis}]
# Code Analysis and LLM Prompt Generation

You are an experienced software engineer using `{language}` programming language skilled in analyzing, summarizing, and writing code. When provided with code, you break it down into its constituent parts, summarize its functionality concisely, and create prompts to guide an LLM in replicating similar outputs.

## Your Tasks:
1. **Code Summary**: Analyze the given code and summarize its purpose, logic, and functionality. Enclose this summary within [SUMMARY] and [/SUMMARY] tags.
2. **Prompt Creation**: Write a clear and specific LLM prompt that, if provided to a language model, would generate code with similar functionality and structure. Enclose the LLM prompt within [LLM_PROMPT] and [/LLM_PROMPT] tags.

Interaction will be in the following way:

### INPUT:
[CODE]
{{code}}
[/CODE]

### OUTPUT:
[SUMMARY]
{{summary}}
[/SUMMARY]

[LLM_PROMPT]
{{prompt}}
[/LLM_PROMPT]
\end{lstlisting}
}

Examples of codes, and corresponding inverse instructions are in \Cref{tab:inst_code_python,tab:inst_code_java,tab:inst_code_go}.

\subsection{\texttt{DroidCollection-Personas} creation}
\label{appx:persona}
To generate \texttt{DroidCollection-Personas}, we started by identifying the main characteristics of a programmer. 
Our final list contains 9 features: Primary Programming Language, Preferred Frameworks, Field of Work, Code Commenting Style, Error-Proneness, Debugging Strategies, Code Aesthetics, Documentation Habits, Function Length Preference. 
The possible values for each feature are listed in \Cref{tab:persona_properties}.

Then we did a Cartesian product to combine all the possible combinations of these properties, and started generating the tasks, which could be performed by this programmer.
For task generation, we used the GPT-4o model, and prompted it in the way shown in \Cref{lst:persona-prompt}. 
{
\lstset{
  basicstyle=\ttfamily,
  frame=single,
  backgroundcolor=\color{gray!5},
  breaklines=true,
}
\begin{lstlisting}[ caption={Prompt for Persona's task generation}, label={lst:persona-prompt}]
I have the following description of a programmer: 
{description}
Write a non-trivial programming task 
which matches what this person probably does at work,
you can ignore some of the person's traits. Return only the task.
\end{lstlisting}
}
After the tasks were generated, we deduplicated them using MinHash with the same parameters as for the dataset filtering. 
After that, the resulting tasks were used for code generation.
\begin{table}[ht]
\centering
\scalebox{0.6}{
\begin{tabular}{@{} l p{5cm} @{}}
\toprule
\textbf{\texttt{Property Name}} & \textbf{\texttt{Values / Options}} \\
\midrule
Primary Programming Language & \makecell[l]{Python, Java, JavaScript, PHP,\\ C, C\#, C++, Go, Ruby, Rust} \\
Field of Work & \makecell[l]{Web Development, AI/ML,  \\ Game Development, \\ System Programming, Embedded \\ Systems, Data Engineering,\\ Research, Distributed \\ Systems Developer, IoT }\\
Code Commenting Style & Concise, Detailed, Minimal \\
Error-Proneness & High, Medium, Low \\
Debugging Strategies & Print Statements, Debugger, Logging \\
Code Aesthetics & Highly Readable, Functional, Minimalist, Hard to Comprehend \\
Documentation Habits & Detailed, Minimal, Occasional \\
Function Length Preference & Short, Medium, Long \\
\bottomrule
\end{tabular}}
\caption{List of attributes and characteristics in \texttt{DroidCollection-Personas}.}
\label{tab:persona_properties}
\end{table}

\subsection{Dataset Statistics}
\label{appx:stats}
In this section, we present key statistics of our dataset and compare them with existing alternatives. 
As shown in \Cref{tab:stats_comparison}, our dataset includes a broader class distribution and shows greater diversity in code structure, as reflected by higher AST depth percentiles and longer line lengths. 
It suggests that our dataset captures more complex and varied code patterns, making it a more challenging and real-life-oriented benchmark for evaluating AI-generated code detection models. The importance of varying code lengths and difficulties is also shown in \Cref{appx:len_score}.
We also show the number of samples per generator, and programming language (not considering the datasets with $<=$ 2 languages or generators). Several qualitative examples of samples belonging to different classes in our dataset are shown in \Cref{tab:code_samples_1,tab:code_samples_2}.



\begin{table*}[ht]
\centering
\scalebox{0.85}{
\begin{tabular}{lcccc}
\toprule
\texttt{\textbf{Metric}} & \texttt{\textbf{CoDet-M4}} & \texttt{\textbf{CodeGPTSensor}} & \texttt{\textbf{GptSniffer}} & \texttt{\textbf{DroidCollection}} \\
\midrule
\texttt{\textbf{AST@75}} & \texttt{\textbf{15.00}} & \texttt{12.00} & \texttt{\textbf{15.00}} & \texttt{\textbf{15.00}} \\
\texttt{\textbf{AST@90}} & \texttt{\textbf{18.00}} & \texttt{15.00} & \texttt{\textbf{18.00}} & \texttt{\textbf{18.00}} \\
\texttt{\textbf{AST@99}} & \texttt{23.00} & \texttt{20.00} & \texttt{23.15} & \texttt{\textbf{25.00}} \\
\texttt{\textbf{Line@75}} & \texttt{90.00} & \texttt{93.00} & \texttt{99.00} & \texttt{\textbf{107.00}} \\
\texttt{\textbf{Line@90}} & \texttt{113.00} & \texttt{112.00} & \texttt{117.00} & \texttt{\textbf{135.00}} \\
\texttt{\textbf{Line@99}} & \texttt{228.00} & \texttt{169.00} & \texttt{153.60} & \texttt{\textbf{314.00}} \\
\midrule
\multirow{4}{*}{\texttt{\textbf{Class Distribution}}} 
& \texttt{AI - 50\%} & \texttt{AI - 50\%} & \texttt{AI - 90\%} & \texttt{AI - 25\%} \\
& \texttt{Human - 50\%} & \texttt{Human - 50\%} & \texttt{Human - 10\%} & \texttt{Human - 47\%} \\
& & & & \texttt{Refined - 13\%} \\
& & & & \texttt{Adv. - 15\%} \\
\midrule
\texttt{\textbf{Avg. \# of samples per language}} & \texttt{\textbf{166,850}} & - & - & \texttt{148,491} \\
\texttt{\textbf{Avg. \# of samples per generator}} & \texttt{\textbf{50,866}} & - & - & \texttt{8,458} \\
\bottomrule
\end
{tabular}}
\caption{Comparison of AST depth percentile, line length percentile, class distribution, and average samples per language/generator between \texttt{DroidCollection} and existing datasets.}
\label{tab:stats_comparison}
\end{table*}

\begin{table}[htbp]
    \centering
\scalebox{0.95}{
    \begin{tabular}{l c c c}
        \toprule
        \texttt{\textbf{Model}} & \multicolumn{3}{c}{\texttt{\textbf{Truncation Length}}}\\
        \cmidrule{2-4}
         & \texttt{\textbf{128}} & \texttt{\textbf{256}} & \textbf{\texttt{512}} \\
        \midrule
        \texttt{GptSniffer} & \texttt{57.05} & \texttt{57.20} & \texttt{56.64} \\
        \texttt{M4} & \texttt{59.69} & \texttt{53.10} & \texttt{51.13} \\
        \texttt{CoDet-M4} & \texttt{72.28} & \texttt{70.62} & \texttt{61.68} \\
        \midrule
        \texttt{DroidDetect-Base} & \texttt{91.90} & \texttt{96.25} & \texttt{99.18} \\
        \texttt{DroidDetect-Large} & \texttt{\textbf{94.91}} & \texttt{\textbf{98.31}} & \texttt{\textbf{99.25}} \\
        \bottomrule
    \end{tabular}
    }
    \caption{Impact of input length truncation (measured using the ModernBERT tokeniser) on weighted F1-scores for binary classification. The most competitive numbers are highlighted in \textbf{bold}.}
    \label{tab:text_length_impact}
\end{table}

\section{Detailed Architectural Ablations}
\label{appx:arch_abl}

\subsection{GCN Experiments}
\label{appx:gnn}
We used a simple 4-layer Graph Convolutional Network (GCN) to evaluate how effectively a GCN can capture structural and semantic features of code.
As input, we utilised AST representations of the code, treating them as graphs. 
To assess the impact of node-level information, we experimented with three types of node features:

\begin{compactitem}
\item \textbf{Dummy features} – no meaningful features were provided at the node level;
\item \textbf{One-hot encoded node types} – encoding the syntactic type of each AST node;
\item \textbf{Node content embeddings} – textual embeddings derived from the string content of each node. 
To reduce computational overhead, we used the \href{https://scikit-learn.org/stable/modules/generated/sklearn.feature_extraction.text.HashingVectorizer.html}{\texttt{HashingVectorizer}}, which converts strings into sparse vectors by hashing tokens to fixed-dimensional indices without maintaining a vocabulary in memory.
\end{compactitem}

As shown in \Cref{tab:node_features}, features based on the textual content of the node yielded the best performance, showing that the semantic information is important in distinguishing between human-written and AI-generated code.

\begin{table}[h]
    \centering
    \scalebox{1.0}{
    \begin{tabular}{lccc}
        \toprule
        \texttt{\textbf{Features}} & \texttt{\textbf{2-class}} & \texttt{\textbf{3-class}} & \texttt{\textbf{4-class}} \\
        \midrule
        \texttt{Dummy} & \texttt{60.02} & \texttt{39.27} & \texttt{34.17} \\
        \texttt{Node Type} & \texttt{50.12} & \texttt{39.54} & \texttt{33.12} \\
        \texttt{Text} & \texttt{\textbf{76.67}} & \texttt{\textbf{59.10}} & \texttt{\textbf{51.14}} \\
        \bottomrule
    \end{tabular}}
    \caption{Comparison of different feature types used as node-level features in a GCN, based on the weighted F1-score on the validation set. The most competitive numbers are highlighted in \textbf{bold}.}
    \label{tab:node_features}
\end{table}

\subsection{CatBoost Experiments}
\label{appx:catboost}

\begin{figure*}[ht]
    \centering
    \includegraphics[width=\linewidth]{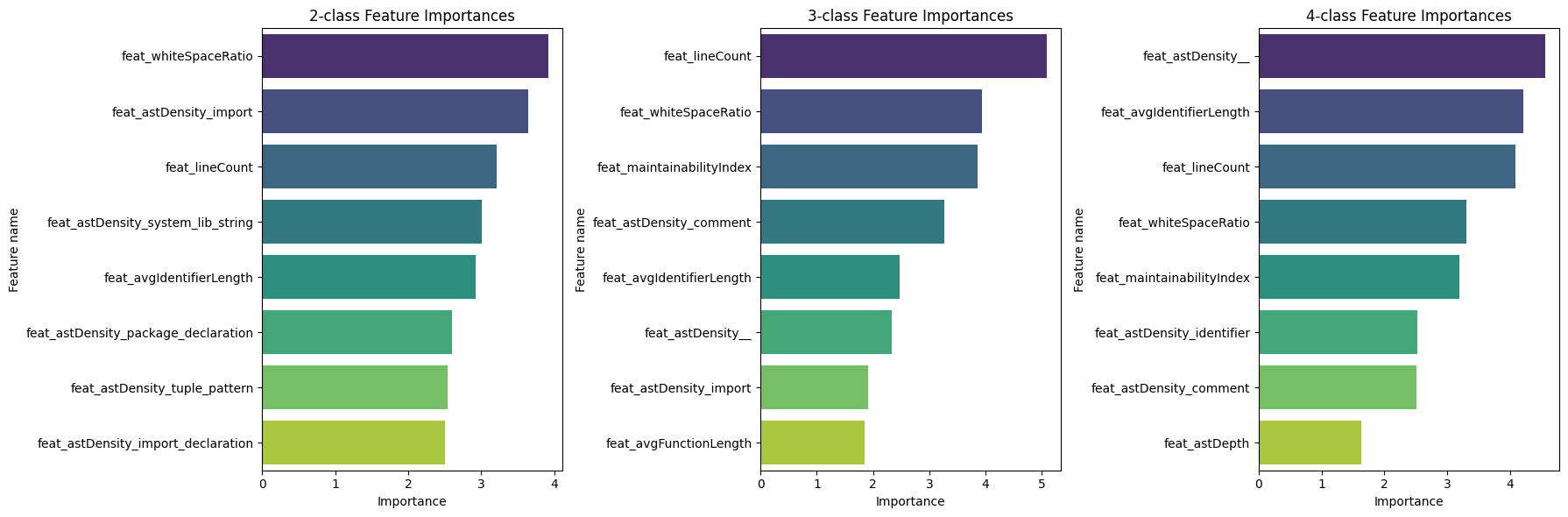}
    \caption{Feature importances}
    \label{fig:cb_features_importancies-plot}
\end{figure*}

Following the experimental methodology of \citet{whodunit} and \citet{orel2025codetm4detectingmachinegeneratedcode}, we computed 733 statistical features that capture various structural properties of code.
These include metrics such as the density of specific AST node types, average line length, whitespace ratio, and the number of empty lines, code maintainability index, among others.

These features were used to train CatBoost classifiers with automatically tuned hyperparameters. \Cref{fig:cb_features_importancies-plot} shows the top unique features ranked by SHAP (SHapley Additive exPlanations) values~\citep{shap}. 
Interestingly, the most informative features vary across the 2-, 3-, and 4-class classification tasks, suggesting that different granularities of classification are dependent on different aspects of code structure.

Nonetheless, some patterns persist across all setups. In particular, features related to the length of identifiers (variable names) and the density of comments consistently present as strong indicators for distinguishing AI-generated/Refined from human-written code. 


\begin{figure}[ht]
    \centering
    \includegraphics[width=\linewidth]{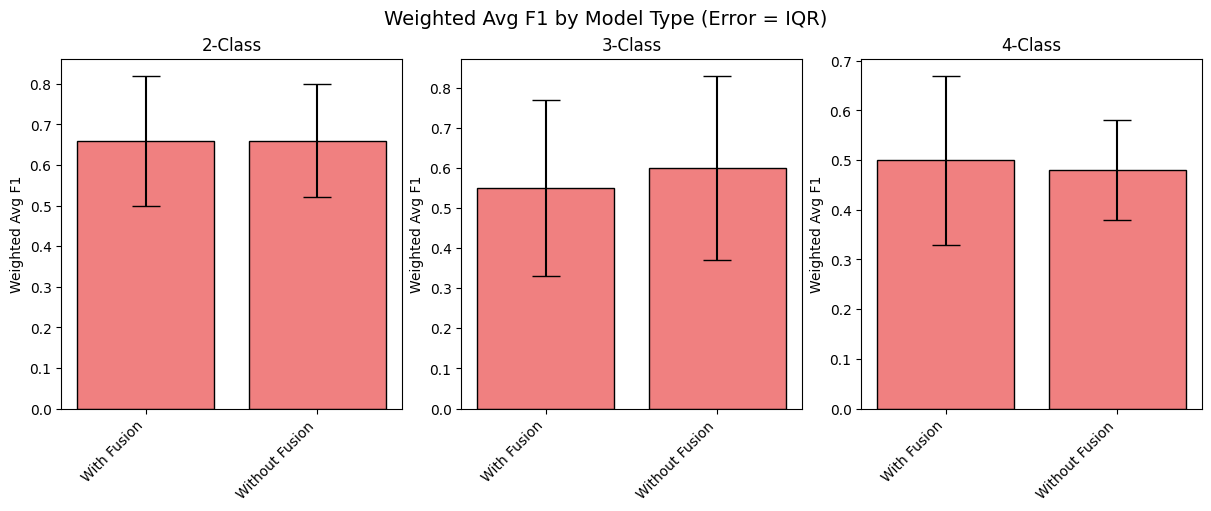}
    \caption{Weighted F1-score comparison between models with and without fusion}
    \label{fig:fusion_barchart}
\end{figure}

\subsection{Does Structure-Based Late-Fusion Improve Robustness?}
\label{appx:fusion_vs_no_fusion}

To decide whether fusion is helpful for improving the detection, we combined the GCN from \Cref{appx:gnn} with our text-only classifier using early fusion of embeddings.
We used OOD-based generalisation, and compared how well the models perform for 2, 3, and 4-class classification in OOD settings (since when trained directly, it is hard to measure the significance of the performance difference), and then compared in which scenarios each method provides a better weighted F1-score. 
\Cref{tab:winrates} shows that there is no clear trend of one approach being better than another: in the binary classification task, there are more ties, fusion has a higher win-rate in 4-class classification, while the model without fusion performs best in the 3-class case.
Then we compared how the difference in F1-scores between models compares to the interquartile range within the model's predictions. As shown in \Cref{fig:fusion_barchart}, the interquartile range is much larger than the model difference, so both models with and without fusion perform nearly equally.
\begin{table}[ht]
\centering
\scalebox{0.85}{
\begin{tabular}{lcccc}
\hline
\multirow{2}{*}{\texttt{\textbf{Classification}}} & \textbf{\texttt{Tie}} & \textbf{\texttt{With}} & \textbf{\texttt{Without}} \\ 
 & \textbf{\texttt{(\%)}} & \textbf{\texttt{Fusion (\%)}} & \textbf{\texttt{Fusion (\%)}} \\
\hline
\texttt{2-Class}     & \texttt{60.0} & \texttt{40.0} &  \texttt{0.0} \\
\texttt{3-Class}     & \texttt{40.0} & \texttt{20.0} & \texttt{40.0} \\
\texttt{4-Class}     & \texttt{20.0} & \texttt{60.0} & \texttt{20.0} \\
\hline
\end{tabular}}
\caption{Comparative task-level win-rates of \texttt{DroidDetect} with and without GCN late-fusion aggregated over OOD classification tasks.}
\label{tab:winrates}
\end{table}

\section{\texttt{DroidDetect} Stress Tests}
\label{appx:stress}

\subsection{Input Length Stress Tests}
\label{appx:len_score}
\Cref{tab:text_length_impact} shows that while other approaches seem to work best on short code snippets, probably because they were trained on shorter code samples (mainly functions) as shown in \Cref{tab:stats_comparison}, our models actually get better with longer code sequences. 
This matters because real code is not usually just a few lines long.
Another important thing is how stable our models remain across different input lengths.
When we cut the input from 512 to 128 tokens,  \texttt{DroidDetect-Base} only drops 7.28 F1-score points (from 99.18 to 91.90), and  \texttt{DroidDetect-Large} drops just 4.34 points (from 99.25 to 94.91). 
This consistency suggests the generalisability of our models to various inputs.

\subsection{Additional OOD Stress Testing}
\label{appx:extra_exps}

To evaluate the generalisation ability of our models, we tested them on additional open-source datasets containing AI-generated code. Specifically, we sampled 15,000 examples from the Swallow-Code dataset~\citep{swallowcode}, a high-quality collection of Python code from The Stack v2~\citep{lozhkov2024starcoder} synthetically refined by LLaMA3.3-70B-Instruct model. This dataset was concurrently released with our work and is highly unlikely to be part of the training distribution of any of our models, thus serving as a strong test for our models' recall on machine-rewritten code.

We also randomly selected 15,000 samples per programming language from The Heap~\citep{heap} dataset. This dataset contains illiberally licensed code with metadata about its presence in existing code-retraining corpora. We specifically filter for samples that are not exact- or near-duplicates with any sample in major pre-training corpora~\citep{li2023starcodersourceyou,lozhkov2024starcoder,DBLP:conf/nips/WeberFAOAALNYAA24}. Jointly, these ensure that our curated split is extremely unlikely to be seen by models during pre-training, thus constituting a stiff test of our detectors' recall on human-written code.

Both \texttt{DroidDetect-Base} and \texttt{DroidDetect-Large} models were tested on these datasets. 
On Swallow-Code, they achieved recall scores of 98.95\% and 99.11\% respectively. On The Heap, \texttt{DroidDetect-Base} reached 94.14\%, while \texttt{DroidDetect-Large} achieved 96.28\%.
It shows that the models trained on our dataset can also work on other datasets robustly.

\section{Error Analysis and Interpretation}
\label{appx:error}

\subsection{Error Analysis}
\blue{In this section we describe and demonstrate common errors, observed in predictions of \texttt{DridDetect} models.}

\paragraph{False Positives and False Negatives}
\blue{Among misclassifications in the binary classification task, we observe that approximately 34\% are false positives, with the remainder being false negatives. This distribution is reasonable given the larger number of negative-class samples in the dataset}

\paragraph{Worst-Performing Language and Domain}
\blue{As shown in \Cref{tab:domain_eval}, Research/DS domain consistently yields the lowest performance for both the Base and the Large \texttt{DroidDetect} models, likely due to the comparatively longer and more complex code typical in this domain.
In \Cref{tab:lang_eval}, we observe that detecting machine-generated code in JavaScript is particularly challenging, which aligns with findings from \citet{orel2025codetm4detectingmachinegeneratedcode}. We attribute this to the greater variability of coding practices found in JavaScript programs - modern frameworks of JavaScript mix coding paradigms (functional programming, OOP) - and it is also not typologically related to the rest of the languages considered in our work.}

\paragraph{Correlation with Preserved Human-Written Code in Hybrid Cases}
\blue{In hybrid scenarios, both models achieve similar F-scores for rewriting and continuation cases (91.42\% and 92.47\% on average). We also find a moderate negative correlation (-0.43 on average) between the proportion of preserved human-written code and the model’s F-score, indicating that the more original human code is retained, the more difficult detection becomes.}

\begin{figure*}[ht]
    \centering
    \begin{minipage}{0.45\textwidth}
        \centering
        \includegraphics[width=\linewidth]{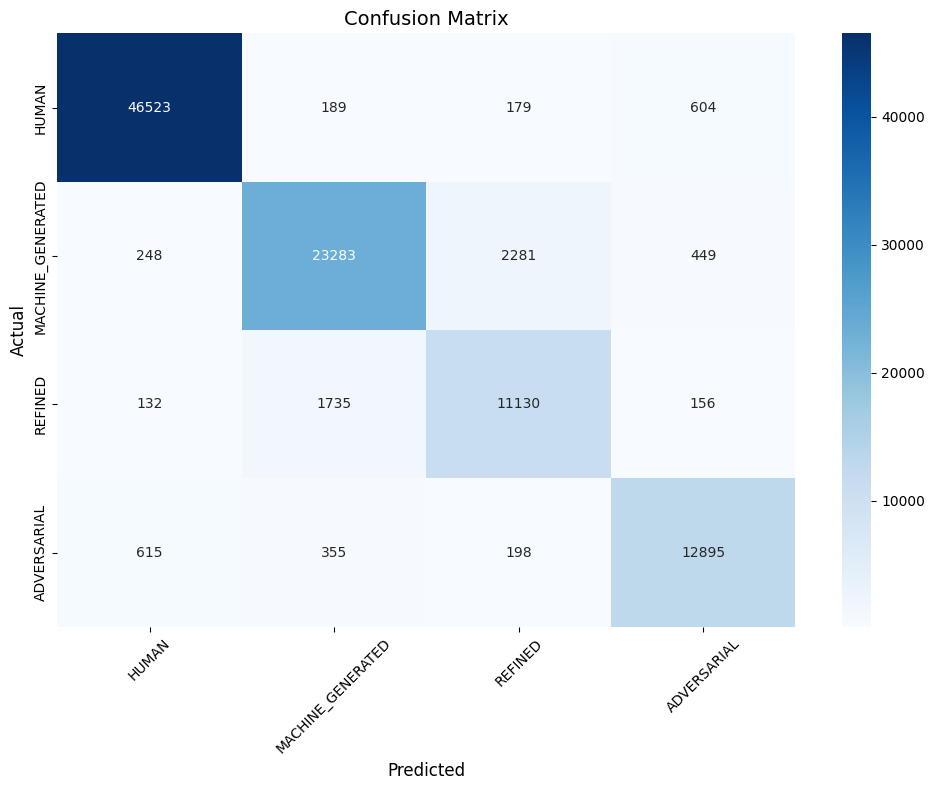} 
        \caption{\texttt{DroidDetect-Base} classifiers confusion matrix}
        \label{fig:base_cm}
    \end{minipage}
    \hfill
    \begin{minipage}{0.45\textwidth}
        \centering
        \includegraphics[width=\linewidth]{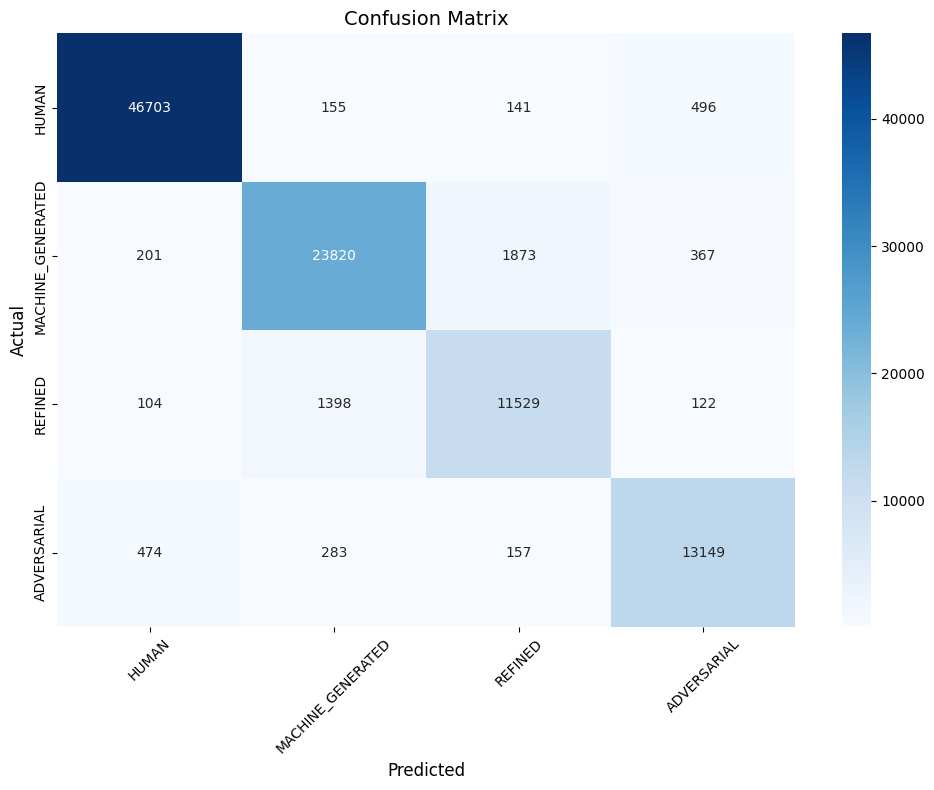} 
        \caption{\texttt{DroidDetect-Large} classifiers confusion matrix}
        \label{fig:large}
    \end{minipage}
    \caption{Confusion Matrixes for \texttt{DroidDetect} models}
    \label{fig:4class_cm}
\end{figure*}

\paragraph{Misclassification on Adversarial Samples}
\blue{As shown in \Cref{fig:4class_cm}, for both the Base and Large \texttt{DroidDetect} models, most misclassifications occur between the AI-generated and Refined classes. Notably, adversarial samples are more frequently misclassified as human-written rather than as other machine-generated (or refined) classes. This suggests that our adversarial training strategy effectively makes these examples more human-like.}

\subsection{Interpretation of Predictions}
\blue{We studied the patterns behind models' predictions, using gradient attribution~\cite{grad_attrib} and the attention maps of our models.
Although these results are somewhat noisy and may not always be fully faithful~\citep{posthoc_dangers, DBLP:journals/coling/LyuAC24}, several common patterns emerge. Notably, much of the model's attention is directed toward punctuation symbols (e.g., colons, semicolons, arrows), consistent with the findings of \citet{DBLP:conf/blackboxnlp/ClarkKLM19}, who showed that BERT-based models frequently focus on punctuation. A more detailed gradient-based analysis reveals that stylistic choices, such as long, descriptive comments and detailed explanations, typical of LLM-generated code, significantly impact the model's decisions. Additionally, consecutive empty lines and extra spaces influence predictions, aligning with the CatBoost feature behaviour described in \Cref{appx:catboost}. Together, these findings suggest that the structural and stylistic conventions of LLM-generated code differ from human-written code.}

\blue{Examples of some misclassifications and their explanations based on gradient attribution and attention maps are given in \Cref{tab:explanations_1,tab:explanations_2}}
\lstdefinestyle{instructionStyle}{
  backgroundcolor=\color{blue!5},
  frame=single,
  breaklines=true,
  keepspaces=true,
  showstringspaces=false
}
\lstdefinestyle{explanationStyle}{
  backgroundcolor=\color{blue!1},
  frame=single,
  breaklines=true,
  keepspaces=true,
  showstringspaces=false
}
\lstdefinestyle{codeStyle}{
  basicstyle=\ttfamily\small,
  frame=single,
  breaklines=true,
  columns=fullflexible,
  keepspaces=true,
  showstringspaces=false,
   keywordstyle=\color{blue},
  commentstyle=\color{gray!60},
  stringstyle=\color{teal},
}
\begin{table*}[!ht]
\centering
\begin{adjustbox}{width=\textwidth}
\begin{tabular}{|p{18cm}|c|c|p{5cm}|}
\hline
\textbf{Code} & \textbf{True Label} & \textbf{Predicted Label} & \textbf{Explanation}  \\
\hline
\begin{lstlisting}[language=Java,style=codeStyle]
import java.io.*;
import java.util.*;
public class randoms {
//just spam uniquely determines LOL. can basically say its 1-D
    public static void main(String[] args) throws IOException {
        BufferedReader br = new BufferedReader(new InputStreamReader(System.in));
        PrintWriter pw = new PrintWriter(System.out);
        StringTokenizer st = new StringTokenizer(br.readLine());
        int n = Integer.parseInt(st.nextToken());
        int k = Integer.parseInt(st.nextToken());
        int[][] nums = new int[n][2];
        for(int i=  0; i<n; i++){
            st = new StringTokenizer(br.readLine());
            nums[i][0] = Integer.parseInt(st.nextToken());
            nums[i][1] = Integer.parseInt(st.nextToken());
        }
        long sum = 0;
        boolean[][] dp = new boolean[n+1][k];
        //a rem uniquely determines brem by overcounting
        //extras accounted for by combinatorics
        dp[0][0] = true;
        for(int i=1; i<=n; i++){
            sum += nums[i-1][0] + nums[i-1][1];
            for(int j = 0; j<k; j++){
                //might be tempted to use int here, but uniquely determined from rem
                int rem = nums[i-1][0]%k;
                dp[i][j] = dp[i-1][(j - rem + k)%k];
                for(int back = 0; back <= Math.min(k-1, nums[i-1][0]); back++){
                    if(nums[i-1][1] + (nums[i-1][0] - back)%k >= k){
                        dp[i][j] = dp[i][j] || dp[i-1][(j - back + k)%k];
                    }
                }
            }
        }
//        System.out.println(Arrays.toString(dp[n]));
        long ret = 0;
        for(int i = 0; i<k; i++){
            if(dp[n][i]){
                ret = (sum - i)/k;
                break;
            }
        }
        pw.println(ret);
        pw.close();
        br.close();
    }
}
\end{lstlisting}
& Human-Written & AI-Generated & 
\begin{lstlisting}[style=explanationStyle]
Gradient attribution shows that this prediction could be based mainly on consequtive whie-spaces in the last comment, use comments that are common for LLMs during code explanation like "basically say it is 1-D", "to use".
\end{lstlisting}
\\
\hline
\begin{lstlisting}[language=C,style=codeStyle]
// Use of this source code is governed by a BSD-style license that can be
// found in the LICENSE file.

#include "chrome/browser/ash/borealis/features.h"

#include <string>

#include "ash/constants/ash_features.h"
#include "base/check.h"
#include "chrome/grit/generated_resources.h"
#include "components/prefs/pref_service.h"

// Generated code

void SetBorealisEnabledAndAllowed(PrefService* pref_service, bool should_allow, bool should_enable) {
  DCHECK(pref_service);
  pref_service->SetBoolean(ash::features::kLauncherShowBorealisAppId, should_allow && should_enable);
  feature_list->InitializeFromCommandLine(kProfileFlag, should_enable);
  cros_settings->SetBoolean(ash::kAccessibilitySpokenFeedbackEnabled, should_enable);
}
\end{lstlisting}
& AI-Generated & Human-Written &
\begin{lstlisting}[style=explanationStyle]
Despite of inclusion of "// Generative code" comment, our models made mistake while classifying this code. Both gradient attribution and attention maps point to the comment which talks about license as the main reason to label it as Human-written code.
\end{lstlisting}
\\
\hline
\end{tabular}
\end{adjustbox}
\caption{Examples of Model misclassification and their explanations (Part 1)}
\label{tab:explanations_1}
\end{table*}

\begin{table*}[!ht]
\centering
\begin{adjustbox}{width=\textwidth}
\begin{tabular}{|p{18cm}|c|c|p{5cm}|}
\hline
\textbf{Code} & \textbf{True Label} & \textbf{Predicted Label} & \textbf{Explanation}  \\
\hline
\begin{lstlisting}[language=C,style=codeStyle]
....
        while (k > 0 and f) {

            if ((v[i] - sum) % k or (v[i] - sum) <= 0) {

                f = 0;

                break;

            }

            ll d = (v[i] - sum) / k;

            sum += (d*2);

            i-=2;

            k -= 2;

        }



        if (f == 1) cout << "YES" << endl;

        else cout << "NO" << endl;

    }
    return 0;
}
\end{lstlisting}
& Human-Written & AI-Generated & 
\begin{lstlisting}[style=explanationStyle]
Extra spacing between lines
\end{lstlisting}
\\
\hline
\begin{lstlisting}[language=Python,style=codeStyle]
def gcdIter(a,b):
	'''
	a, b: positive integers
	
	returns: a positive integer, the greatest common divisor of a & b.
	'''
	# Your code here
	test=min(a,b)
	while test>=1:
		if a%test==0 and b%test==0:
			return test
		test-=1
	   

import fractions

S,tc=list(map(int,input().split(' ')))
#print S,tc
already_occured=[]
while tc:
	tc=tc-1
	A=int(input())
	#gcd=gcdIter(A,S)
	gcd=fractions.gcd(A,S)
	if gcd in already_occured:
		print(-1)
	else:
		already_occured.append(gcd)
		print(gcd)

\end{lstlisting}
& Human-Written & AI-Generated &
\begin{lstlisting}[style=explanationStyle]
Both attention maps and gradient attribution trigger on "# Your code here" comment, which is similar to "Here is your code", which is a common part of LLMs response.
\end{lstlisting}
\\
\hline
\begin{lstlisting}[language=C,style=codeStyle]
#include <tuple>

template <int Arg>
class TestObj
{
public:
  int getArg()
  {
    return Arg;
  }
};

//----------------------------------------------------------------------
// Define a template class that we can specialize with an enumeration
//----------------------------------------------------------------------
enum class EnumType
{
    Member,
    Subclass
};

template <EnumType Arg> class EnumTemplate;
                                          
//----------------------------------------------------------------------
// Specialization for use when "Arg" is "EnumType::Member"
//----------------------------------------------------------------------
template <>
class EnumTemplate<EnumType::Member> 
{
public:
    EnumTemplate(int m) :
        m_member(m)
    {
    }

    int getMember() const
    {
        return m_member;
    }

protected:
    int m_member;
};
\end{lstlisting}
& AI-Generated & Human-Written &
\begin{lstlisting}[style=explanationStyle]
Consequitive spaces and the format of comments, highlighted by attention maps and gradient attribution.
\end{lstlisting}
\\
\hline
\end{tabular}
\end{adjustbox}
\caption{Examples of Model misclassification and their explanations (Part 2)}
\label{tab:explanations_2}
\end{table*}

\section{Qualitative Examples}
\label{appx:qual}

\subsection{Inverse Instructions Examples}
\label{appx:inv_samples}
In \Cref{tab:inst_code_python,tab:inst_code_java,tab:inst_code_go} we show examples of code with the corresponding inverse instructions.

\begin{table*}[!ht]
\centering
\begin{adjustbox}{width=\textwidth}
\begin{tabular}{|c|p{7cm}|p{18cm}|}
\hline
\textbf{Language} & \textbf{Instruction} & \textbf{Code} \\
\hline
C++ & 
\begin{lstlisting}[style=instructionStyle]
Write a C-style function to retrieve the bus speed of a specific device, given its UART's baud clock divisor, base baud clock, and time base reading functionality. The function should set the DLAB, read the DLL and DLM, calculate the baud clock divisor, update the DLAB, and then loop to find the first edge of the baud clock, get the starting and ending time base values, and finally compute the bus speed from these values.
\end{lstlisting}
& 
\begin{lstlisting}[language=C,style=codeStyle]
static unsigned long __init
prpmc750_get_bus_speed(void)
{
unsigned long tbl_start, tbl_end;
unsigned long current_state, old_state, bus_speed;
unsigned char lcr, dll, dlm;
int baud_divisor, count;

/* Read the UART's baud clock divisor */
lcr = readb(PRPMC750_SERIAL_0_LCR);
writeb(lcr | UART_LCR_DLAB, PRPMC750_SERIAL_0_LCR);
dll = readb(PRPMC750_SERIAL_0_DLL);
dlm = readb(PRPMC750_SERIAL_0_DLM);
writeb(lcr & ~UART_LCR_DLAB, PRPMC750_SERIAL_0_LCR);
baud_divisor = (dlm << 8) | dll;

/*
* Use the baud clock divisor and base baud clock
* to determine the baud rate and use that as
* the number of baud clock edges we use for
* the time base sample. Make it half the baud
* rate.
*/
count = PRPMC750_BASE_BAUD / (baud_divisor * 16);

/* Find the first edge of the baud clock */
old_state = readb(PRPMC750_STATUS_REG) & PRPMC750_BAUDOUT_MASK;
do {
current_state = readb(PRPMC750_STATUS_REG) &
PRPMC750_BAUDOUT_MASK;
} while(old_state == current_state);

old_state = current_state;

/* Get the starting time base value */
tbl_start = get_tbl();

/*
* Loop until we have found a number of edges equal
* to half the count (half the baud rate)
*/
do {
do {
current_state = readb(PRPMC750_STATUS_REG) &
PRPMC750_BAUDOUT_MASK;
} while(old_state == current_state);
old_state = current_state;
} while (--count);

/* Get the ending time base value */
tbl_end = get_tbl();

/* Compute bus speed */
bus_speed = (tbl_end-tbl_start)*128;

return bus_speed;
}
\end{lstlisting}
\\
\hline
Python & 
\begin{lstlisting}[style=instructionStyle]
Generate Python code that performs data reading, trimming, and datetime conversion for a set of files located in a specified directory. The files should be named in a pattern `met_sum_insitu_1_obop_hour_20<Year>.txt`, and the data should be trimmed to remove irrelevant columns, handle missing values, and create a datetime column using the existing year, month, day, and hour columns. The resulting data should be returned as a processed pandas DataFrame
\end{lstlisting}
& 
\begin{lstlisting}[language=Python,style=codeStyle]
def metTrim():
# ---- initial reading of data
root = r'C:\Users\ARL\Desktop\MetData'
ext = list(range(12, 20)) # yearly extensions

colnames = ['na', 'yr', 'mo', 'dy', 'hr', 'dir', 'spd', 'steady', 'na', 'na', 'na', 'na', 'na', 'na']
met = pd.DataFrame(columns=colnames) # preallocate df
for yr in ext:
# read in data
data = pd.read_csv(root + r'\met_sum_insitu_1_obop_hour_20{}.txt'.format(yr), delim_whitespace=True,
header=None)
data.columns = colnames # apply col names
met = met.append(data) # append to list
print('Data Imported')

# ---- trimming data
met = met.drop('na', axis=1) # drop na cols
met = met.replace(-999.9, np.nan) # turn missing val to nan
met = met.replace(-9, np.nan)
met = met.replace(-999, np.nan)
met = met.replace(-99.9, np.nan)
met = met.dropna(axis=0, how='any') # remove rows with nan vals

# ---- convert date to datetime
metInt = met.applymap(int) # make sure values are ints
dates = createDatetime(metInt['yr'].values,
metInt['mo'].values,
metInt['dy'].values,
metInt['hr'].values)

met['datetime'] = dates # add it as a new column
met = met.drop(['yr', 'mo', 'dy', 'hr'], axis=1) # drop old date columns

return met
\end{lstlisting}\\
\hline
\end{tabular}
\end{adjustbox}
\caption{Code and Corresponding Inverse Instructions (C++, Python)}
\label{tab:inst_code_python}
\end{table*}
\begin{table*}[!ht]
\centering
\begin{adjustbox}{width=\textwidth}
\begin{tabular}{|c|p{7cm}|p{18cm}|}
\hline
\textbf{Language} & \textbf{Instruction} & \textbf{Code} \\
\hline
Java & 
\begin{lstlisting}[style=instructionStyle]
Given a file with blog entries and their respective dates, write a Java method to read the file, sort the entries in reverse chronological order, and separate them into lists for published and unpublished entries. The method should handle varying file existence conditions and log errors accordingly. Assume the blog entries are stored in lines of the file, with each entry consisting of a date in 'yyyyMMdd' format followed by the entry's ID.
\end{lstlisting}
& 
\begin{lstlisting}[language=Java,style=codeStyle]
private void readIndex(boolean published) {
File indexFile;
if (published) {
indexFile = new File(blog.getIndexesDirectory(), "blogentries-published.index");
} else {
indexFile = new File(blog.getIndexesDirectory(), "blogentries-unpublished.index");
}

if (indexFile.exists()) {
try {
BufferedReader reader = new BufferedReader(new FileReader(indexFile));
String indexEntry = reader.readLine();
while (indexEntry != null) {
indexEntries.add(indexEntry);

// and add it to the internal memory structures
Date date = new Date(Long.parseLong(indexEntry));
Day day = blog.getBlogForDay(date);

if (published) {
publishedIndexEntries.add(indexEntry);
day.addPublishedBlogEntry(indexEntry);
} else {
unpublishedIndexEntries.add(indexEntry);
day.addUnpublishedBlogEntry(indexEntry);
}

indexEntry = reader.readLine();
}

reader.close();
} catch (Exception e) {
log.error("Error while reading index", e);
}
}

Collections.sort(indexEntries, new ReverseBlogEntryIdComparator());
Collections.sort(publishedIndexEntries, new ReverseBlogEntryIdComparator());
Collections.sort(unpublishedIndexEntries, new ReverseBlogEntryIdComparator());
}
\end{lstlisting}
\\
\hline
JavaScript & 
\begin{lstlisting}[style=instructionStyle]
Generate a JavaScript function named Teth, which is a constructor function, outside of its parent function. Teth should inherit all properties and methods from its parent function and add a new method - getChain. This method should return the string "teth". The Teth function should attach its prototype to its parent's prototype, moving it one level away. The code should utilize the Node.js environment and should include documentation to describe the purpose of the function and its newly added method.
\end{lstlisting}
& 
\begin{lstlisting}[language=Java,style=codeStyle]
var Btc = require('./btc');
var bitcoin = require('bitcoinjs-lib');
var _ = require('lodash');

var Tbtc = function() {
// this function is called externally from BaseCoin
// replace the BaseCoin prototype with the local override prototype, which inherits from BaseCoin
// effectively, move the BaseCoin prototype one level away
this.__proto__ = Tbtc.prototype;
this.network = bitcoin.networks.testnet;
};

Tbtc.prototype.__proto__ = Btc.prototype;

Tbtc.prototype.getChain = function() {
return 'tbtc';
};

module.exports = Tbtc;
\end{lstlisting}\\
\hline
\end{tabular}
\end{adjustbox}
\caption{Code and Corresponding Inverse Instructions (Java, JavaScript)}
\label{tab:inst_code_java}
\end{table*}
\begin{table*}[!ht]
\centering
\begin{adjustbox}{width=\textwidth}
\begin{tabular}{|c|p{7cm}|p{18cm}|}
\hline
\textbf{Language} & \textbf{Instruction} & \textbf{Code} \\
\hline
Go & 
\begin{lstlisting}[style=instructionStyle]
Write a Go function that creates a set of constant labels for Prometheus metrics based on ISI cluster settings.
The function should check if a site has been specified for the ISI cluster and create labels with both the cluster name and site name if available.
Otherwise, it should create labels with only the cluster name.
Include a log statement to display the created labels.
\end{lstlisting}
& 
\begin{lstlisting}[language=Go,style=codeStyle]
func CreateConstLabels() error {
//Only create a const label for site if a site has been specified.
if IsiCluster.Site != "" {
ConstLabels = prometheus.Labels{"cluster": IsiCluster.Name, "site": IsiCluster.Site}
} else {
ConstLabels = prometheus.Labels{"cluster": IsiCluster.Name}
}
log.Debugf("ConstLables are %v", ConstLabels)
return nil
}
\end{lstlisting}
\\
\hline
C\# & 
\begin{lstlisting}[style=instructionStyle]
Generate C# code that prompts the user for a string and then checks if it is a palindrome by comparing characters from the start and end of the string, moving towards the center. If the characters match, print a message indicating they are equal; otherwise, print a message indicating they are not equal. The code should handle strings of any length and input user characters until the string length is reached.
\end{lstlisting}
& 
\begin{lstlisting}[language=C,style=codeStyle]
using System;

namespace _76
{
class Program
{
static void Main(string[] args)
{
int cont = 0;

Console.WriteLine("Quantos caracteres  seu palindromo tem?");
int max = int.Parse(Console.ReadLine());

string [] caractere = new string [max];

for (cont = 0; cont < max; cont++)
{
Console.WriteLine("Digite" + (cont + 1) + "  caracteres da palavra");
caractere[cont] = Console.ReadLine();
}

int inverso = cont;

for (cont = 0; cont < max; cont++)
{
if (caractere[cont] == caractere[inverso])
{
Console.WriteLine("O " + (cont + 1) + " caractere  " + (inverso + 1) + " caractere");
}
Console.WriteLine("O " + (cont + 1) + " caractere " + (inverso + 1) + " caractere ");
inverso--;
}
}
}
}
\end{lstlisting}\\
\hline
\end{tabular}
\end{adjustbox}
\caption{Code and Corresponding Inverse Instructions (Go, C\#)}
\label{tab:inst_code_go}
\end{table*}

\subsection{Dataset Samples}
\label{appx:samples}
In this appendix (\Cref{tab:code_samples_1,tab:code_samples_2}), we provide a small portion of code per class, written by different models in different languages. 
To check the diversity of our dataset, it is suggested to check the release repository.
\lstdefinestyle{codeStyle}{
  basicstyle=\ttfamily,
  frame=single,
  breaklines=true,
  columns=fullflexible,
  keepspaces=true,
  showstringspaces=false,
   keywordstyle=\color{blue},
  commentstyle=\color{gray!60},
  stringstyle=\color{teal},
}

\begin{table*}[!ht]
\centering
\begin{adjustbox}{width=\textwidth}
\begin{tabular}{|c|c|c|p{18cm}|}
\hline
\textbf{Language} & \textbf{Class} & \textbf{Generator} & \textbf{Code} \\
\hline
Python & \makecell{LLM-Refined \\(re-written)} & Qwen2.5-72B & 
\begin{lstlisting}[language=Python,style=codeStyle]
from collections import defaultdict
class Solution:
MAXPRIME = 100001
isPrime = [0] * (MAXPRIME + 1)
isPrime[0] = isPrime[1] = -1
def __init__(self):
for i in range(2, MAXPRIME):
if isPrime[i] == 0:
isPrime[i] = i
for multiple in range(i * i, MAXPRIME + 1, i):
if isPrime[multiple] == 0:
isPrime[multiple] = i
def largestComponentSize(self, A):
label = defaultdict(int)
roots = {}
def find_root(key):
if key not in roots:
roots[key] = key
if roots[key]!= key:
roots[key] = find_root(roots[key])
return roots[key]
def merge_roots(k1, k2):
r1, r2 = find_root(k1), find_root(k2)
if r1!= r2:
r1, r2 = min(r1, r2), max(r1, r2)
label[r1] += label[r2]
roots[r2] = r1
return r1
for x in A:
root_id = None
prime_factors = set()
while self.isPrime[x]!= -1:
p = self.isPrime[x]
root_id = find_root(p) if root_id is None else merge_roots(root_id, p)
x //= p
label[root_id] -= 1
return -min(label.values())
\end{lstlisting}
\\
\hline
C & \makecell{Human-written} & Human & 
\begin{lstlisting}[language=C,style=codeStyle]
int
pack_white(char *ptr)
{
int cnt = 0;
char *tptr, ch;
if (ptr == NULL)
return (0);
tptr = ptr;
while (isspace(*tptr))
tptr++;
for (;;) {
while ((ch = *tptr) != '\0' && !isspace(ch)) {
*ptr++ = ch;
tptr++;
}
while (isspace(*tptr))
tptr++;
if (*tptr == '\0')
break;
*ptr++ = ' ';
cnt++;
}
*ptr = '\0';
return (cnt);
}
\end{lstlisting}
\\
\hline
\end{tabular}
\end{adjustbox}
\caption{Code samples by different models (Part 1)}
\label{tab:code_samples_1}
\end{table*}
\begin{table*}[!ht]
\centering
\begin{adjustbox}{width=\textwidth}
\begin{tabular}{|c|c|c|p{18cm}|}
\hline
\textbf{Language} & \textbf{Class} & \textbf{Generator} & \textbf{Code} \\
\hline
Go & \makecell{Adversarial} & Qwen2.5-Coder-7B-Instruct & 
\begin{lstlisting}[language=Go,style=codeStyle]
package main
import (
"context"
"fmt"
"golang.org/x/sync/errgroup"
)
func parallelExecute(ctx context.Context, runners ...func(context.Context) error) error {
g := new(errgroup.Group)
for _, runner := range runners {
runner := runner
g.Go(func() error {
return runner(ctx)
})
}
return g.Wait()
}
func main() {
ctx := context.Background()
runners := []func(context.Context) error{
func(ctx context.Context) error {
select {
case <-ctx.Done():
return ctx.Err()
case <-time.After(2 * time.Second):
return fmt.Errorf("runner 1 error")
}
},
func(ctx context.Context) error {
select {
case <-ctx.Done():
return ctx.Err()
case <-time.After(3 * time.Second):
return fmt.Errorf("runner 2 error")
}
},
}
err := parallelExecute(ctx, runners...)
if err != nil {
fmt.Println("First error encountered:", err)
}
}
\end{lstlisting}
\\
\hline
JavaScript & \makecell{AI-Generated} & Yi-Coder-9B & 
\begin{lstlisting}[language=Java,style=codeStyle]
class Vector2D {
#x;
#y;
constructor(x, y) {
this.#x = x;
this.#y = y;
}
setX(x) {
this.#x = x;
}
setY(y) {
this.#y = y;
}
getX() {
return this.#x;
}
getY() {
return this.#y;
}
add(vector) {
this.#x += vector.getX();
this.#y += vector.getY();
return this;
}
compare(vector) {
return this.#x === vector.getX() && this.#y === vector.getY();
}
}
\end{lstlisting}
\\ \hline
\end{tabular}
\end{adjustbox}
\caption{Code samples by different models (Part 2)}
\label{tab:code_samples_2}
\end{table*}

\end{document}